\documentclass[prd,aps,nofootinbib,tightenlines]{revtex4}

\newcommand{\checked}[1]{}

\usepackage{extarrows}
\usepackage{amsmath}
\usepackage{tabularx}
\usepackage{amsmath}
\usepackage{graphicx}
\usepackage{booktabs}
\usepackage{color}

\newcommand{\be}{\begin{equation}}
\newcommand{\ee}{\end{equation}}
\newcommand{\ba}{\begin{eqnarray}}
\newcommand{\ea}{\end{eqnarray}}
\newcommand{\Tr}{{\rm Tr\,}}
\newcommand{\la}{\label}

\begin{document}

\title {Resummed Gluon Propagator and Debye Screening Effect in a Holonomous Plasma}

\author{Yun Guo$^{a,b}$ and Zhenpeng Kuang$^{a}$}
\affiliation{
$^a$ Department of Physics, Guangxi Normal University, Guilin, 541004, China\\
$^b$ Guangxi Key Laboratory of Nuclear Physics and Technology, Guilin, 541004, China\\}

\begin{abstract}

Based on the Dyson-Schwinger equation, we compute the resummed gluon propagator in a holonomous plasma that is described by introducing a constant background field for the vector potential $A_{0}$. Due to the transversality of the holonomous Hard-Thermal-Loop in gluon self-energy, the resummed propagator has a similar Lorentz structure as that in the perturbative Quark-Gluon Plasma where the holonomy vanishes. As for the color structures, since diagonal gluons are mixed in the over-complete double line basis, only the propagators for off-diagonal gluons can be obtained unambiguously. On the other hand, multiplied by a projection operator, the propagators for diagonal gluons, which exhibit a highly non-trivial dependence on the background field, are uniquely determined after summing over the color indices. As an application of these results, we consider the Debye screening effect on the in-medium binding of quarkonium states by analyzing the static limit of the resummed gluon propagator. In general, introducing non-zero holonomy merely amounts to modifications on the perturbative screening mass $m_D$ and the resulting heavy-quark potential, which remains the standard Debye screened form, is always deeper than the screened potential in the perturbative Quark-Gluon Plasma. Therefore, a weaker screening, thus a more tightly bounded quarkonium state can be expected in a holonomous plasma. In addition, both the diagonal and off-diagonal gluons become distinguishable by their modified screening masses ${\cal M}_D$ and the temperature dependence of the ratio ${\cal M}_D/T$ shows a very similar behavior as that found in lattice simulations.

\end{abstract}

\maketitle

\section{Introduction}\la{intro}

At high temperatures, the properties of the Quark-Gluon Plasma (QGP) created during the ultra-relativistic heavy-ion collisions can be computed in the Hard-Thermal-Loop (HTL) resummed perturbation theory.  On the other hand, at low temperatures, the confined phase can be modeled by a hadron resonance gas. The challenge appears in the intermediate region, termed as ``semi"-QGP where neither of the above mentioned theoretical tools is reliable since the effects of non-perturbative physics play an important role.

As the order parameter for deconfinement in $SU(N)$ gauge theory, the values of Polyakov loop are non-zero but less than unity in ``semi"-QGP. The partial deconfinement is described by introducing non-zero holonomy for Polyakov loops. To do so, one can consider a classical background field $A_0^{\rm cl}$ as a diagonal and traceless color matrix for the time-like component of the vector potential. Thermodynamics of a holonomous plasma can be analyzed by computing the effective potential in the (constant) background field $A_0^{\rm cl}$ which takes the eigenvalues of the thermal Wilson line as variables\cite{Belyaev:1991gh,Bhattacharya:1992qb,KorthalsAltes:1993ca,KorthalsAltes:1999cp,Dumitru:2013xna,Reinosa:2015gxn,Maelger:2017amh,Guo:2018scp}. Perturbatively, the effective potential reaches a minimum when the background field vanishes. Therefore, a complete deconfinement happens at all temperatures. In order to drive the transition to confinement, non-perturbative terms, which generate complete eigenvalue repulsion in the confining phase, have to be included. Constructed in such a way, the matrix models have been widely studied in recent years, not only for pure gauge theories, but also for Quantum ChromoDynamics (QCD) with dynamical quarks\cite{Meisinger:2001cq,Dumitru:2010mj,Dumitru:2012fw,Guo:2014zra,Pisarski:2016ixt}.

The physics in ``semi"-QGP is of particular interest because the temperatures probed in most of the high energy experiments carried out at Relativistic Heavy Ion Collider (RHIC) and the Large Hadron Collider (LHC) are not far above the critical temperature. Besides the thermodynamical properties, physical quantities near thermal equilibrium have also been investigated with nonzero holonomy for Polyakov loops which exhibit different behaviors as compared to those in the perturbative QGP. For example, the shear viscosity computed in ``semi"-QGP is suppressed near the critical temperature\cite{Hidaka:2009ma}. As ideal electromagnetic signals, the production of dileptons calculated in a matrix model has a mild enhancement, conversely, the production of photons is strongly suppressed in the presence of a background field\cite{Gale:2014dfa,Hidaka:2015ima}. In addition, discussions on the transport coefficients as well as the collisional energy loss of a heavy quark in ``semi"-QGP can be found in Refs.~\cite{Singh:2018wps,Singh:2019cwi,Lin:2013efa}.

For processes involving soft momentum exchange, it is necessary to use the resummed propagator which includes arbitrary number of self-energy insertions into the bare propagator. The expression of the resummed gluon propagator is well-known in the perturbative QGP where the holonomy vanishes. As an important quantity in many theoretical and phenomenological applications, viscous corrections to the resummed gluon propagator in thermal equilibrium has been studied in Refs.~\cite{Dumitru:2007hy,Burnier:2009yu,Dumitru:2009fy,Du:2016wdx,Nopoush:2017zbu}. Furthermore, influence due to the presence of a magnetic field has also been considered in recent works\cite{Hattori:2016idp,Bandyopadhyay:2016fyd,Singh:2017nfa}.  On the other hand, in a holonomous plasma,
the explicit form of the resummed gluon propagator with $A_{0}^{\rm cl}\neq 0$ adopted in the foresaid works relies on certain approximations, for example, one needs to assume infinitely large number of the colors or neglect an anomalous term $\sim T^3$ in the perturbative gluon self-energy that only appears with non-vanishing holonomy.

There is a long history of the computation of the gluon self-energy in a holonomous plasma. In Ref.~\cite{Hidaka:2009hs}, it has been computed at one-loop order in the HTL perturbation theory and there is a non-transverse piece showing up in the obtained result. As argued in Ref.~\cite{KorthalsAltes:2020ryu}, gauge invariant sources, which are nonlinear in the gauge potential $A_0$, give rise to a novel constrained contribution at one-loop order which restores the transversality of the holonomous gluon self-energy. 
As already mentioned before, perturbatively, the system would be always in a completely deconfined vacuum since the equations of motion lead to a vanishing background field. In order to generate a non-zero holonomy dynamically, an effective theory has been proposed in Ref.~\cite{Hidaka:2020aa} where additional contributions from two dimensional ghosts were introduced into the action and the resulting gluon self-energy remains transverse. Given the holonomous gluons self-energy, the main obstacle to compute the resummed propagator lies in the complicated color structure when one performs the inversion through the Dyson-Schwinger equation. In this paper, we make a first attempt to calculate the resummed gluon propagator in ``semi"-QGP for general $SU(N)$. In addition, as a direct application of the obtained results, we also consider the modifications on the screening masses due to a non-vanishing holonomy which provides important information on the in-medium binding of quarkonium states.

The rest of the paper is organized as the following. In section \ref{review}, we briefly introduce the double line basis which will be adopted in our calculation. In section \ref{ba}, the bare gluon propagator in a holonomous plasma denoted as $(D_0)_{\mu\nu}^{ab,cd}(P^{ab})$ is discussed. It is an intuitive example to understand the complicated color structure we will encounter in the computation of the resummed propagator. For completeness, in section \ref{se}, we give a short review on the holonomous gluon self-energy obtained in previous studies. Based on the Dyson-Schwinger equation, the resummed gluon propagator ${\tilde{D}}_{\mu\nu}^{ab,cd}(P^{ab})$ is computed in section \ref{resum} where the calculations are carried out for the diagonal and off-diagonal gluons separately. 
After analytically continued to real time, in section \ref{screening}, the static limit of the propagator ${\tilde{D}}_{\mu\nu}^{ab,cd}(\omega, {\bf p})$ is analyzed which gives new insights into the screening effect in a holonomous plasma.  A short summary can be found in section \ref{summary}. In addition, some details about the calculations performed in this work are provided in three appendices.

\section{The inverse propagators at tree level in the double line basis}\la{review}

In the presence of a constant background field $A^{{\rm cl}}_0$, the double line basis has been widely used in previous studies to compute the effective potential\cite{Guo:2018scp} as well as the quark/gluon self-energies for $SU(N)$ gauge theories\cite{Hidaka:2009hs}. For completeness, we will briefly review the double line basis and give the inverse propagators at tree level for later use. More details can be found in Ref.~\cite{Cvitanovic:1976am}.

The generators of the fundamental representation are given by the projection operators,
\be
\la{21}
(t^{ab})_{cd} = \frac{1}{\sqrt{2}}{\cal P}^{ab}_{cd}\, ,
\ee
with
\be\la{22}
{\cal P}^{ab}_{cd} = \delta^{a}_{c}\delta^{b}_{d}-\frac{1}{N}\delta^{ab}\delta_{cd}\, .
\ee
For $SU(N)$, these color indices $a,b,c$ and $d$ run from $1$ to $N$. The $N^2-N$ off-diagonal generators with $a\neq b$ are normalized as
\be
\textrm{tr}(t^{ab}t^{ba})= \frac{1}{2}\, .
\ee
In addition, we have $N$ diagonal generators $t^{aa}$ which satisfy
\be
\textrm{tr}(t^{aa}t^{bb})= \frac{1}{2}\bigg(\delta^{ab}-\frac{1}{N}\bigg)\, .
\ee
In the above equations, $a$ and $b$ are fixed indices and there is no summation over them. In the double line basis, the number of generators for $SU(N)$ is $N^2$, therefore, this basis is over-complete.

Notice that the upper indices $ab$ of the generators refer to the indices in the adjoint representation which are denoted by a pair of the fundamental indices. The lower indices $cd$ refer to the matrix elements in the fundamental representation.

The Lagrangian of $SU(N)$ gauge theory is given by
\be
{\cal L}=\frac{1}{2} {\rm {tr}}(G_{\mu\nu}^2)\quad\quad {\rm and}\quad \quad G_{\mu\nu}=\partial _\mu A_\nu-\partial _\nu A_\mu-i[A_\mu,A_\nu]\, .
\ee
The gauge fields can be expanded around some fixed classical values as $A_\mu=A^{\rm cl}_\mu + g B_\mu$ and $B_\mu$ corresponds to the quantum fluctuation. Including the gauge fixing term (the gauge fixing parameter is denoted by $\xi$) and ghost fields $\eta$ in the Lagrangian
\be
{\cal L}_{\rm gauge}=\frac{1} {\xi} \Tr(D_{\mu}^{{\rm cl}}B_\mu)^2- 2  \Tr(\bar{\eta} D_{\mu}^{{\rm cl}}D_{\mu}\eta)\,,
\ee
we can write down the corresponding terms related to the (inverse) gluon propagator at tree level in the action, ${\cal S} = \int d^4 x \,  {\cal L}$, as the following
\be
{\cal S}= \int d^4  x \, \Tr \Big\{B^{\mu}\Big(-(D_{\rho}^{\rm {cl}})^2\delta_{\mu\nu}+(1-\frac{1}{\xi})D_{\mu}^{\rm {cl}}D_{\nu}^{\rm {cl}}+2 i g [G^{\rm cl}_{\mu\nu},\,\cdots]\Big)B^{\nu}\Big\}+\cdots\, .
\ee
In the above equations, the classical covariant derivative is defined as $D_{\mu}^{\rm {cl}}=\partial _\mu-i g [A_{\mu}^{\rm {cl}}, \, \cdots]$.

We consider the classical background field as a constant diagonal matrix for the time-like component of the vector potential, namely, $(A^{\rm cl}_0)_{ab}=Q^a \delta_{ab}$ with $\sum_{a=1}^{N} Q^a=0$ for $SU(N)$ gauge group.
Consequently, the classical covariant derivative acting upon the fields in the adjoint representation has a simple form in momentum space, $D_\mu^{{\rm cl}} t^{ab} \rightarrow -i P_\mu^{ab} t^{ab}$ and the corresponding momentum associated with an adjoint color index $ab$ reads
\ba
\la{28}
P^{ab}_\mu =(p^{ab}_0,\textbf{p})=(\omega_{n}+Q^a-Q^b,\textbf{p})\, ,
\ea
where $\omega_{n}$ is the Matsubara frequencies of bosons. Then it is straightforward to write down the inverse bare gluon propagator in momentum space
\be\la{inbare1}
(D_0^{-1})^{ab,cd}_{\mu\nu}(P^{ab})=\frac{\delta {\cal S}}{\delta B_\mu^{ba}(P) \delta B_\nu^{dc}(-P)}=\Big( (P^{ab})^2\delta_{\mu\nu}-(1-\frac{1}{\xi})P_\mu^{ab}P_\nu^{ab}\Big){\cal P}^{ab,cd}\, .
\ee
As one can see, this is a trivial generalization of $(D_0^{-1})^{ab,cd}_{\mu\nu}(P)$ in the case where $A_0^{{\rm cl}}=0$ since there is only a constant and color-dependent shift in the energies.

The inverse ghost propagator can be obtained in the same way and the result is given by
\be
\frac{\delta {\cal S}}{\delta \eta^{ba}(P) \delta {\bar {\eta}}^{dc}(-P)}= (P^{ab})^2 {\cal P}^{ab,cd}\, .
\ee
Adding the quark contribution $\bar{\psi} (\displaystyle{\not} D+m) \psi$ to the pure gauge action, the inverse quark propagator has the following explicit form\footnote{In the fundamental representation, the classical covariant derivative $D_{\mu}^{\rm {cl}}=\partial _\mu-i g A_{\mu}^{\rm {cl}}$ acting upon the fermionic field $\psi^a$ is $D_{\mu}^{\rm {cl}} \psi^a(x) \rightarrow - i P^a_\mu \psi^a(P)$.}
\be
\frac{\delta {\cal S}}{\delta \psi^{a}(P) \delta \bar{\psi}^{b}(-P)}=(-i \displaystyle{\not} P^a+m)\delta^{ab}\, ,
\ee
where $P^a_\mu$ associated with a fundamental color index $a$ is defined as $P^a_\mu  =  (p^a_{0},\textbf{p})=(\tilde{\omega}_{n}+Q^a,\textbf{p})$ with $\tilde{\omega}_{n}$ being the Matsubara frequencies of fermions.

The inverse quark propagator has a trivial color structure $\delta^{ab}$, therefore, the corresponding bare propagator is a diagonal matrix in color space, explicitly, we have
\be\la{baqk}
\left \langle \psi^{a}(P) \bar{\psi}^{b}(-P)\right \rangle =\frac{\delta^{ab}}{-i \displaystyle{\not} P^a+m}\, .
\ee
On the other hand, the inverse gluon/ghost propagator containing the projection operator ${\cal P}^{ab,cd}$ and the color structure of the bare propagator is not as simple as the quark propagator. We will give a detailed discussion in the next section.

\section{The bare gluon propagator in a constant background field}\la{ba}

Before we start to discuss the resummed gluon propagator, it is worthwhile to analyze the color structure of the bare propagator in the double line basis. As we will see, there exists an issue that the exact forms of the gluon propagators can not be uniquely determined. For general $SU(N)$ gauge theories, the $N^2-N$ off-diagonal generators $t^{ab}$ with $a \neq b$ are identical to those in the Carton space, however, as compared to the $N-1$ diagonal generators in the Carton space, the $N$ diagonal generators $t^{aa}$ are over-complete which is believed to be the original of such an ambiguity we will encounter. One thing to note is that although the discussion on the bare gluon propagator turns to be simple, the very similar strategy can be generalized to compute the resummed propagator which will be considered in Sec.~\ref{resum}.

We rewrite the inverse bare propagator as given in Eq.~(\ref{inbare1}) as
\be
(D_0^{-1})^{ab,cd}_{\mu\nu}(P^{ab})=\bigg[(P^{ab})^2\bigg(\delta_{\mu\nu}-\frac{P_\mu^{ab} P_{\nu}^{ab}}{(P^{ab})^2}\bigg)+\frac{1}{\xi}P_\mu^{ab} P_{\nu}^{ab}\bigg]{\cal P}^{ab,cd}\, ,
\ee
where the mutually orthogonal projections $\delta_{\mu\nu}-(P_\mu^{ab} P_{\nu}^{ab})/(P^{ab})^2$ and $P_\mu^{ab} P_{\nu}^{ab}$ are the natural extension of those used in $A_0^{\rm cl}=0$. The unity in the Lorentz space is defined as $\delta_{\mu\nu}$ while in the color space it is given by ${\cal P}^{ab,cd}$. Therefore, the gluon propagator $D^{ab,cd}_{\mu\nu}$ (either the bare propagator $(D_0)^{ab,cd}_{\mu\nu}$ or the resummed one ${\tilde D}^{ab,cd}_{\mu\nu}$) satisfies the following identity
\be\la{id}
\sum_{\sigma,ef}(D^{-1})^{ab,ef}_{\mu\sigma}(P^{ab})\cdot D^{fe,cd}_{\sigma\nu}(P^{fe})=\sum_{\sigma,ef}D^{ab,ef}_{\mu\sigma}(P^{ab})\cdot (D^{-1})^{fe,cd}_{\sigma\nu}(P^{fe})=\delta_{\mu\nu}{\cal P}^{ab,cd}\, .
\ee

The bare gluon propagator $(D_0)^{ab,cd}_{\mu\nu}(P^{ab})$ used in Ref.~\cite{Hidaka:2009hs} reads
\be\la{bapro1}
\bigg[\delta_{\mu\nu}-(1-\xi)\frac{P_\mu^{ab} P_{\nu}^{ab}}{(P^{ab})^2}\bigg]\frac{1}{(P^{ab})^2}{\cal P}^{ab,cd}\, ,
\ee
and one can easily check the above form of the propagator satisfies the desired identity, Eq.~(\ref{id}). However, as we will show, Eq.~(\ref{bapro1}) is not a unique solution and the bare propagator is not necessary to be proportional to the projection operator ${\cal P}^{ab,cd}$. More generally, we assume the following form for the bare propagator
\be
(D_0)^{ab,cd}_{\mu\nu}(P^{ab})={\cal {X}}_0^{ab,cd}\bigg(\delta_{\mu\nu}-\frac{P_\mu^{ab} P_{\nu}^{ab}}{(P^{ab})^2}\bigg)+{\cal {Z}}_0^{ab,cd}P_\mu^{ab} P_{\nu}^{ab}\, .
\ee
We first consider the bare propagator for off-diagonal gluons which is denoted as $(D_0)^{ab,cd}_{\mu\nu}$ with $a\neq b$. By definition, we can show that
\ba\la{did1}
&&\sum_{ef,\sigma}(D^{-1}_0)_{\mu\sigma}^{ab,ef}(P^{ab})\cdot (D_0)_{\sigma\nu}^{fe,cd}(P^{fe})\xlongequal[]{a\neq b}\sum_{\sigma}(D_0^{-1})_{\mu\sigma}^{ab,ba}(P^{ab})
(D_0)_{\sigma\nu}^{ab,cd}(P^{ab})\, ,\nonumber \\
&=&(P^{ab})^2{\cal {X}}_0^{ab,cd}\bigg(\delta_{\mu\nu}-\frac{P_\mu^{ab} P_{\nu}^{ab}}{(P^{ab})^2}\bigg)+\frac{1}{\xi}{\cal {Z}}_0^{ab,cd}P_\mu^{ab} P_{\nu}^{ab}={\cal P}^{ab,cd}\delta_{\mu\nu}\xlongequal[]{a\neq b}\delta^{ad}\delta^{bc}\delta_{\mu\nu}\, .
\ea
This equation holds when the following conditions are satisfied
\begin{equation}
\left\{
\begin{aligned}
&(P^{ab})^2{\cal {X}}_0^{ab,cd} =\delta^{ad}\delta^{bc} \\
&-\delta^{ad}\delta^{bc}\frac{1}{(P^{ab})^2}+\frac{1}{\xi}{\cal {Z}}_0^{ab,cd}(P^{ab})^2 =0 \\
\end{aligned}
\right.
\end{equation}
This leads to the results ${\cal {X}}_0^{ab,cd} =\delta^{ad}\delta^{bc}/(P^{ab})^2$ and ${\cal {Z}}_0^{ab,cd}=\xi\delta^{ad}\delta^{bc}/(P^{ab})^4$ for $a \neq b$. Therefore, we find
\be\la{bareoff}
(D_0)^{ab,cd}_{\mu\nu}(P^{ab})\xlongequal[]{a\neq b}\delta^{ad}\delta^{bc}\bigg\{\frac{1}{(P^{ab})^2}\bigg(\delta_{\mu\nu}-\frac{P_\mu^{ab} P_{\nu}^{ab}}{(P^{ab})^2}\bigg)+\frac{\xi}{(P^{ab})^4}P_\mu^{ab} P_{\nu}^{ab}\bigg\}\, .
\ee
Alternatively, one can consider $\sum_{\sigma,ef}(D_0)^{ab,ef}_{\mu\sigma}(P^{ab})\cdot (D^{-1}_0)^{fe,cd}_{\sigma\nu}(P^{fe})=\delta_{\mu\nu}{\cal P}^{ab,cd}$ with $c\neq d$. Consequently, $(D_0)^{ab,cd}_{\mu\nu}(P^{ab})$ with $c\neq d$ can be determined which is identical to Eq.~(\ref{bareoff}) as expected. In addition, the above result also indicates vanishing gluon propagators $(D_0)_{\mu\nu}^{aa,cd}(P)$ and $(D_0)_{\mu\nu}^{cd,aa}(P)$ when $c\neq d$.

As we can see, the bare propagators for off-diagonal gluons can be uniquely determined which have relatively simple color structures proportional to the projection operator ${\cal P}^{ab,cd}$. The only non-vanishing component $(D_0)^{ab,ba}_{\mu\nu}(P^{ab})$ for $a\neq b$ is the same as the one used in Ref.~\cite{Hidaka:2009hs}, see Eq.~(\ref{bapro1}).

Next, we consider the bare propagators for diagonal gluons, $(D_0)^{aa,cc}_{\mu\nu}\equiv (D_0)^{a,c}_{\mu\nu}$. From here on, the diagonal color index $aa$ will be denoted by a single letter $a$ in order to keep the notation compact. Similarly, we have
\ba
\sum_{ef,\sigma}(D^{-1}_0)_{\mu\sigma}^{a,ef}(P)\cdot (D_0)_{\sigma\nu}^{fe,c}(P^{fe})&=&\bigg(P^2{\cal {X}}_0^{a,c}-\frac{P^2}{N}\sum_e{\cal {X}}_0^{e,c}\bigg)\bigg(\delta_{\mu\nu}-\frac{P_\mu P_{\nu}}{P^2}\bigg)\nonumber \\
&+&\bigg(\frac{P^2}{\xi}{\cal {Z}}_0^{a,c}-\frac{P^2}{N\xi}\sum_e{\cal {Z}}_0^{e,c}\bigg)P_\mu P_{\nu}
={\cal P}^{a,c}\delta_{\mu\nu}\, .\nonumber \\
\ea
It leads to the following equations
\begin{equation}
\left\{
\begin{aligned}
&P^2{\cal {X}}_0^{a,c}-\frac{P^2}{N}\sum_e{\cal {X}}_0^{e,c} ={\cal P}^{a,c} \\
&-\frac{1}{P^2}{\cal P}^{a,c}+\bigg(\frac{P^2}{\xi}{\cal {Z}}_0^{a,c}-\frac{P^2}{N\xi}\sum_e{\cal {Z}}_0^{e,c}\bigg) =0 \\
\end{aligned}
\right.
\end{equation}
We start with the first equation for ${\cal {X}}_0^{a,c}$. For a given $c$, there are $N$ equations correspond to $a=1,2,\cdots, N$. However, they are not completely independent. Because $\sum_a{\cal P}^{a,cd}=0$, any equation can be derived from the other $N-1$ equations. Therefore, we drop the equation with $a=c$ and the other $N-1$ independent equations for ${\cal {X}}_0^{a,c}$ can be written as
\be\la{eqd}
(1-\frac{1}{N})({\cal {X}}_0^{a,c}-{\cal {X}}_0^{c,c})-\frac{1}{N}\sum_{e\neq a,c}({\cal {X}}_0^{e,c}-{\cal {X}}_0^{c,c})=-\frac{1}{NP^2}\, , \quad a=1,\cdots,c-1,c+1,\cdots,N\,.
\ee
Instead of ${\cal {X}}_0^{a,c}$, we consider ${\cal {X}}_0^{a,c}-{\cal {X}}_0^{c,c}$ with $a\neq c$. For a fixed $c$, there are $N-1$ unknowns in Eq.~(\ref{eqd}). The coefficient matrix reads
\begin{equation}\la{cm}
\begin{pmatrix}
1-\frac{1}{N} & -\frac{1}{N}  & \cdots   & -\frac{1}{N}   \\
-\frac{1}{N} & 1-\frac{1}{N}  & \cdots   & -\frac{1}{N}  \\
\vdots & \vdots  & \ddots   & \vdots  \\
-\frac{1}{N} & -\frac{1}{N}  & \cdots  & 1-\frac{1}{N}
\end{pmatrix}
\, ,
\end{equation}
which is a $(N-1)\times(N-1)$ symmetric matrix with $1-\frac{1}{N}$ being the diagonal elements and $-\frac{1}{N}$ for other elements. The determinant of this matrix can be calculated for general $SU(N)$ and the result simply equals $\frac{1}{N}$. Using Cramer's rule, it is straightforward to obtain the solutions of Eq.~(\ref{eqd}) and the $N^2-N$ unknowns ${\cal {X}}_0^{a,c}-{\cal {X}}_0^{c,c}$ are all equal to $-1/P^2$ where $a,c=1,2,\cdots,N$ and $c\neq a$.

Similarly, the $N-1$ independent equations for ${\cal {Z}}_0^{a,c}-{\cal {Z}}_0^{c,c}$ are given by
\be\la{eqz}
(1-\frac{1}{N})({\cal {Z}}_0^{a,c}-{\cal {Z}}_0^{c,c})-\frac{1}{N}\sum_{e\neq a,c}({\cal {Z}}_0^{e,c}-{\cal {Z}}_0^{c,c})=-\frac{\xi}{N P^4}\, , \quad a=1,\cdots,c-1,c+1,\cdots,N\,,
\ee
which suggests a simple relation ${\cal {Z}}_0^{a,c}-{\cal {Z}}_0^{c,c}=\frac{\xi}{P^2}({\cal {X}}_0^{a,c}-{\cal {X}}_0^{c,c})$.  As a result, we have
\be\label{ds1}
\left\{
\begin{aligned}
&{\cal {X}}_0^{a,c}={\cal {X}}_0^{c,c}-\frac{1}{P^2} \\
&{\cal {Z}}_0^{a,c}={\cal {Z}}_0^{c,c}-\frac{\xi}{P^4} \\
\end{aligned}
\right.\,,\quad a\neq c\, .
\ee

On the other hand, we should also consider
\be\la{ddi2}
\sum_{ef,\sigma}(D_0)_{\mu\sigma}^{a,ef}(P)\cdot (D^{-1}_0)_{\sigma\nu}^{fe,c}(P^{fe})={\cal P}^{a,c}\delta_{\mu\nu}\, ,
\ee
which determines the unknowns ${\cal {X}}_0^{a,c}-{\cal {X}}_0^{a,a}$ and ${\cal {Z}}_0^{a,c}-{\cal {Z}}_0^{a,a}$. Following exactly the same procedure as the above, we can further show that
\be\label{ds2}
\left\{
\begin{aligned}
&{\cal {X}}_0^{a,c}={\cal {X}}_0^{a,a}-\frac{1}{P^2} \\
&{\cal {Z}}_0^{a,c}={\cal {Z}}_0^{a,a}-\frac{\xi}{P^4} \\
\end{aligned}
\right.\,,\quad a\neq c\, .
\ee
Summing up Eqs.~(\ref{ds1}) and (\ref{ds2}), the relations among the unknowns in the propagator are found as
\begin{equation}\label{ds3}
\left\{
\begin{aligned}
&{\cal {X}}_0^{c,a}={\cal {X}}_0^{a,c}={\cal {X}}_0^{a,a}-\frac{1}{P^2}={\cal {X}}_0^{c,c}-\frac{1}{P^2} \\
&{\cal {Z}}_0^{c,a}={\cal {Z}}_0^{a,c}={\cal {Z}}_0^{a,a}-\frac{\xi}{P^4}={\cal {Z}}_0^{c,c}-\frac{\xi}{P^4} \\
\end{aligned}
\right.\,,\quad a\neq c\, .
\end{equation}

The above result shows that the bare propagators $(D_0)_{\mu\nu}^{a,c}(P)$ can be uniquely determined as long as any one component (such as $(D_0)_{\mu\nu}^{1,1}(P)$ or $(D_0)_{\mu\nu}^{1,2}(P)$) is specified. In general, we can only get the following expression
\be\la{gepro}
(D_0)_{\mu\nu}^{a,a}(P)-(D_0)_{\mu\nu}^{c,a}(P)=(D_0)_{\mu\nu}^{a,a}(P)-(D_0)_{\mu\nu}^{a,c}(P)=\bigg[\delta_{\mu\nu}-(1-\xi)\frac{P_\mu P_{\nu}}{P^2}\bigg]\frac{1}{P^2}\, .
\ee
Eq.~(\ref{gepro}) is familiar which is the same as the bare propagators for the $N^2-1$ gluons when the standard choice for the generators of a gauge group is adopted\footnote{In this case, the propagator is proportional to $\delta^{AB}$ where $A$ and $B$ refer to adjoint indices running from $1$ to $N^2-1$ for $SU(N)$. }.  Furthermore, it is interesting to point out that if one special constraint $\sum_a(D_0)_{\mu\nu}^{a,c}(P)=0$ is imposed, the diagonal gluon propagators can be uniquely determined which are identical to those in Eq.~(\ref{bapro1}).

Although the exact form of the gluon propagator for diagonal gluons can not be specified without extra constraint, one can still draw a conclusion based on the above obtained results. Like the inverse bare gluon propagator, $(D_0)_{\mu\nu}^{ab,cd}(P^{ab})$ is also symmetric when we flip the Lorentz and/or color indices, namely $(D_0)_{\mu\nu}^{ab,cd}(P^{ab})=(D_0)_{\nu\mu}^{dc,ba}(P^{dc})$. As a consequence, $\sum_{\sigma,ef}(D_0^{-1})^{ab,ef}_{\mu\sigma}\cdot (D_0)^{fe,cd}_{\sigma\nu}=\delta_{\mu\nu}{\cal P}^{ab,cd}$ will automatically lead to $\sum_{\sigma,ef}(D_0)^{ab,ef}_{\mu\sigma}\cdot (D_0^{-1})^{fe,cd}_{\sigma\nu}=\delta_{\mu\nu}{\cal P}^{ab,cd}$ and vice versa.

On the other hand, the ambiguity of the bare gluon propagator $(D_0)_{\mu\nu}^{a,c}(P)$ doesn't turn to be a real problem. Generally speaking, not the individual components of the propagator, but some proper combinations of them are of practical interest. For example, the quantity $\sum_{abcd}{\cal P}_{ab,cd}D_{\mu\nu}^{ab,cd}(P^{ab})$ is the one we are interested in which will be studied in Sec.~\ref{screening}. With the above results, we can show that
\ba\la{uni}
&&\sum_{abcd}{\cal P}_{ab,cd}(D_0)_{\mu\nu}^{ab,cd}(P^{ab})=\sum_{ab}(D_0)_{\mu\nu}^{ab,ba}(P^{ab})-\frac{1}{N}\sum_{ac}(D_0)_{\mu\nu}^{aa,cc}(P)\nonumber\\
&&={\sum_{ab}}^{\prime}(D_0)_{\mu\nu}^{ab,ba}(P^{ab})-\frac{1}{N}{\sum_{ac}}^{\prime}(D_0)_{\mu\nu}^{aa,cc}(P)+(1-\frac{1}{N})\sum_{a}(D_0)_{\mu\nu}^{aa,aa}(P)\nonumber\\
&&={\sum_{ab}}^{\prime}(D_0)_{\mu\nu}^{ab,ba}(P^{ab})+(N-1)\bigg[\delta_{\mu\nu}-(1-\xi)\frac{P_\mu P_{\nu}}{P^2}\bigg]\frac{1}{P^2}\, .
\ea
In the above equation, we have used Eqs.~(\ref{ds1}) and (\ref{ds2}). $(D_0)_{\mu\nu}^{ab,ba}(P^{ab})$ for $a\neq b$ can be found in Eq.~(\ref{bareoff}) and ${\sum}^{\prime}$ indicates terms with $a=b$ are excluded. The previous ambiguity doesn't show up after summing over the color indices. In Eq.~(\ref{uni}), contributions from the diagonal gluons don't affected by the background filed while for the off-diagonal gluons there is a simple shift in the energies, {\em i.e.}, $p_0\rightarrow p_0^{ab}$. For vanishing background filed, the above result reduces to the following expected from
\ba
\sum_{abcd}{\cal P}_{ab,cd}(D_0)_{\mu\nu}^{ab,cd}(P)=(N^2-1)\bigg[\delta_{\mu\nu}-(1-\xi)\frac{P_\mu P_{\nu}}{P^2}\bigg]\frac{1}{P^2}\, .
\ea

Finally, we would like to mention that the bare ghost propagator shares similar properties as the gluon propagator. The corresponding discussion can be carried out by using exactly the same method as above.

\section{The gluon self-energy at non-zero holonomy}\la{se}

In this work, we are interested in the resummed gluon propagators $\tilde{D}_{\mu\nu}^{ab,cd}(P^{ab})$ which is expected to provide information on the screening effects induced by the light partons in a holonomous plasma. Given the discussions on the bare propagators in Sec.~\ref{ba}, one may naturally conjecture that the individual components of the resummed gluon propagators for diagonal gluons can not be uniquely determined, however, the ambiguity would be absent in some special combinations, such as $\sum_{abcd}{\cal P}_{ab,cd}{\tilde D}_{\mu\nu}^{ab,cd}(P^{ab})$. In addition, how the resummed gluon propagator in the perturbative QGP would be modified by the background field $A_0^{{\rm cl}}$ in a holonomous plasma is obviously another interesting question that needs to be addressed.

The above issues can be clarified by computing the resummed gluon propagator ${\tilde D}_{\mu\nu}^{ab,cd}(P^{ab})$ based on the Dyson-Schwinger equation where the holonomous gluon self-energy $\Pi_{\mu\nu}^{ab,cd}(P^{ab})$ needs to be inserted. Within the perturbation theory, the leading order $\Pi_{\mu\nu}^{ab,cd}(P^{ab})$ has been calculated in Ref.~\cite{Hidaka:2009hs} within HTL approximation where Eq.~(\ref{bapro1}) was used for the bare propagators. In principle, one can choose other possible forms for the diagonal propagators, however, the obtained gluon self-energy doesn't depend on any specific choice. This is easy to see by considering the bare propagator and its associated structure constant $f^{ab,cd,ef}=i(\delta^{ad}\delta^{cf}\delta^{eb}-\delta^{af}\delta^{cb}\delta^{ed})/\sqrt{2}$. For  the contribution from the gluon-loop diagram, we have the color summation $\sum_{ab}f^{ef,gh,ab}D_{\mu\nu}^{ab,cd}(P^{ab})$ for each bare gluon propagator. Here, only diagonal components matter, therefore, we need to show
\ba
\sum_{a}f^{ef,gh,aa}{(D_0)}_{\mu\nu}^{a,c}(P)&=&f^{ef,gh,cc}{(D_0)}_{\mu\nu}^{c,c}(P)+\sum_{a\neq c}f^{ef,gh,aa}{(D_0)}_{\mu\nu}^{a,c}(P)\nonumber\\
&=&\bigg[\delta_{\mu\nu}-(1-\xi)\frac{P_\mu P_{\nu}}{P^2}\bigg]\frac{1}{P^2}f^{ef,gh,cc}+\sum_{a}f^{ef,gh,aa}{(D_0)}_{\mu\nu}^{d,c}(P)\nonumber\\
&=&\bigg[\delta_{\mu\nu}-(1-\xi)\frac{P_\mu P_{\nu}}{P^2}\bigg]\frac{1}{P^2}f^{ef,gh,cc}\,.
\ea
In the above equation, all the color indices except $a$ are fixed and $c\neq d$ applies. In the second line of this equation, we have used the relation given in Eq.~(\ref{gepro}). Clearly, there is no ambiguity appearing after summing over the color indices. The same analysis also applies to the ghost-loop diagram. In addition, such an issue doesn't show up in the bare quark propagator, therefore, the perturbative gluon self-energy has been uniquely determined although a specified form of the gluon/ghost bare propagator was used in Ref.~\cite{Hidaka:2009hs}.

For completeness, we also list the explicit result of the HTL gluon self-energy in a constant background field which reads\footnote{in Ref.~\cite{Hidaka:2009hs}, contributions from fermions are also obtained. We drop the fermionic terms for simplicity.}
\ba
\label{pipt}
 \Pi_{{\rm pert};\mu\nu}^{ab,cd}(P^{ab})&=&\left({\cal K}^{ab, cd}(q)+\left(m_{\rm gl}^2\right)^{ab,cd}(q)\right)\left(\Pi_T(P^{ab})A_{\mu\nu}(P^{ab})+ \Pi_L(P^{ab})B_{\mu\nu}(P^{ab})\right)\nonumber\\
&-&{\cal K}^{ab, cd}(q)M_\mu M_\nu\, ,
\ea
where
\ba
{\cal K}^{ab,cd}(q)
= - \frac{4 \pi}{3} \frac{g^2 T^3}{p_0^{ab}} \delta^{ad} \delta^{bc}
\sum_{e = 1}^{N} \left( B_3(q^{ae}) + B_3(q^{eb}) \right)
\label{gluon_selfJ}
\ea
and
\ba
\left(m_{\rm gl}^2\right)^{ab,cd}(q)
= g^2 T^2 \left[
\delta^{ad} \delta^{bc} \sum_{e = 1}^{N}\left( B_2(q^{ae}) + B_2(q^{eb})\right)
-2 \delta^{ab} \delta^{cd} B_2(q^{ac})\right]
\, .
\label{gluon_selfH}
\ea
In the above equations, $q^{ab}\equiv q^a-q^b$ and $q^a=Q^a/(2\pi T)$. In addition, we also use $q$ to denote any arbitrary $q^a$ for $a=1,2,\cdots ,N$. The Bernoulli polynomials $B_n(x)$ are periodic functions of $x$, with period $1$. For $0\le x \le 1$, the first four Bernoulli polynomials as relevant in the present work take the following forms
\be\la{aa3}
B_1(x)=x-\frac{1}{2}\,,\quad B_2(x)=x^2-x+\frac{1}{6}\, ,\quad B_3(x)=x^3-{\frac{3}{2}}x^2+\frac{1}{2}x\,,\quad B_4(x)=x^4-2x^3+x^2-\frac{1}{30}\, .
\ee
For arbitrary values of $x$, the argument of the Bernoulli polynomials should be understood as $x-[x]$ with $[x]$ the largest integer less than $x$, which is nothing but the modulo function.

For convenience, the gluon self-energy in Eq.~(\ref{pipt}) has been expressed in terms of the mutually orthogonal projection operators which are defined as
\ba
A_{\mu\nu}(P^{ab})&=&\delta_{\mu\nu}-\frac{P_\mu^{ab}P_\nu^{ab}}{(P^{ab})^2}-\frac{\tilde{M}_\mu^{ab}\tilde{M}_\nu^{ab}}{(\tilde{M}^{ab})^2}\, ,\nonumber\\
B_{\mu\nu}(P^{ab})&=&\frac{(P^{ab})^2}{(M \cdot P^{ab})^2}\frac{{\tilde{M}_\mu^{ab}\tilde{M}_\nu^{ab}}}{(\tilde{M}^{ab})^2}\, .
\ea
Here, $M_{\mu}$ is the heat-bath vector, which in the local rest frame is given by $M_{\mu}=(1,0,0,0)$ and
\ba
\tilde{M}_\mu^{ab}=M_\mu-\frac{M \cdot P^{ab}}{(P^{ab})^2}P_\mu^{ab}\, ,
\ea
is the part that is orthogonal to $P_\mu^{ab}$. In addition, the two structure functions $\Pi_T(P^{ab})$ and $\Pi_L(P^{ab})$ take similar forms as their counterparts in $A_0^{\rm cl}=0$,
\ba
\Pi_T(P^{ab})&=&\frac{(i p_0^{ab})^2}{2p^2}\left[1-\frac{(i p_0^{ab})^2-p^2}{2i p_0^{ab} p}\ln \frac{i p_0^{ab}+p}{i p_0^{ab}-p}\right]\, ,\nonumber\\
\Pi_L(P^{ab})&=&\frac{(i p_0^{ab})^2}{p^2}\left[\frac{i p_0^{ab}}{2p}\ln \frac{i p_0^{ab}+p}{i p_0^{ab}-p}-1\right]\, ,
\ea
where $p=|{\bf p}|$.

For vanishing background field, the new hard thermal loop ${\cal K}^{ab,cd}$ doesn't contribute while Eq.~(\ref{gluon_selfH}) becomes the perturbative Debye mass square $m^2_D=N g^2 T^2/3$ times the projection operator ${\cal P}^{ab,cd}$. As a result, the gluon self-energy reduces to the following well-known result\footnote{Like the bare gluon propagator, if the standard choice for the generators is used, we will have $\delta^{AB}$ instead of ${\cal P}^{ab,cd}$ in Eq.~(\ref{pizero}). In this case, the gluon self-energy is a diagonal matrix in color space with equal non-zero elements.},
\ba
\label{pizero}
\Pi_{{\rm pert};\mu\nu}^{ab,cd}(P, q=0)=m_D^2 \Big[\Pi_T(P)A_{\mu\nu}(P)+ \Pi_L(P)B_{\mu\nu}(P)\Big]{\cal P}^{ab,cd} \, .
\ea

Although having the expected form at vanishing holonomy, the holonomous gluon self-energy as given in Eq.~(\ref{pipt}) is not transverse, $P_{\mu}^{ab}\Pi_{{\rm pert};\mu\nu}^{ab,cd}(P^{ab})=-p_0^{ab} {\cal K}^{ab, cd}(q) M_\nu$. 
On the other hand, as discussed in Ref.~\cite{KorthalsAltes:2020ryu}, gauge invariant sources, which are nonlinear in the gauge potential $A_0$, generate a novel constrained contribution $\sim {\cal K}^{ab, cd}(q)M_\mu M_\nu$ to the gluon self-energy at one-loop order in the perturbation theory. It exactly cancels the last term in Eq.~(\ref{pipt}) and the total gluon self-energy $\Pi_{{\rm cons};\mu\nu}^{ab,cd}(P^{ab})$ remains transverse,
\ba
\label{pito}
\Pi_{{\rm cons};\mu\nu}^{ab,cd}(P^{ab})&=&\left({\cal K}^{ab, cd}(q)+\left(m_{\rm gl}^2\right)^{ab,cd}(q)\right)\left(\Pi_T(P^{ab})A_{\mu\nu}(P^{ab})+ \Pi_L(P^{ab})B_{\mu\nu}(P^{ab})\right)\, .\nonumber\\
\ea
However, for any gauge invariant source, there is an unexpected discontinuity in the free energy appearing at order $\sim g^3$ as the holonomy vanishes. For details, see Refs.~\cite{KorthalsAltes:2020ryu, KorthalsAltes:2019yih}. 

It turns out that both the non-transversality and discontinuity as mentioned above are related to the anomalous term $\sim {\cal K}^{ab, cd}(q)$ involving the third Bernoulli polynomial $B_3(q)$ in the holonomous gluon self-energy. Adding the constraint contribution only leads to a partial cancellation of ${\cal K}^{ab, cd}(q)$. 
Another issue arising here is the non-vanishing expectation value of the holonomous color current which indicates that extra term should be included in the action to ensure a vanishing result.
 As discussed in Ref.~\cite{Hidaka:2020aa}, embedding two dimensional ghosts isotropically into four dimensions, a new contribution proportional to the second Bernoulli polynomial $B_2(q)$ appears in the effective potential which modifies the equations of motion and leads to a nonzero holonomy $q\sim C/T^2$ at high temperature. Here, the cutoff scale $C$ has dimensions of mass square and corresponds to the upper limit of the transverse momentum $k_\perp ^2$ of the embedded fields. In such an effective theory, the holonomous color current vanishes as expected because the contribution from two dimensional ghosts exactly cancels that from perturbative theory. Furthermore, the free energy to $\sim g^3$ becomes continuous due to the absence of the anomalous term ${\cal K}^{ab, cd}(q)$ in the holonomous gluons self-energy which finally takes the following simple form
\ba
\label{pito}
\Pi_{{\rm eff};\mu\nu}^{ab,cd}(P^{ab})&=&\left(\left(m_{\rm gl}^2\right)^{ab,cd}(q)+g^2 C\frac{N}{4\pi^2}{\cal P}^{ab,cd}\right)\left(\Pi_T(P^{ab})A_{\mu\nu}(P^{ab})+ \Pi_L(P^{ab})B_{\mu\nu}(P^{ab})\right)\, .\nonumber\\
\ea
Since the two projection operators $A_{\mu\nu}(P^{ab})$ and $B_{\mu\nu}(P^{ab})$ are both orthogonal to $P_\mu^{ab}$, the gluon self-energy from the effective theory is also transverse.

Finally, it is worth to note that the above gluon self-energies are all symmetric under the exchange of the Lorentz indices $\mu \leftrightarrow\nu$ as well as the color indices $a \leftrightarrow d$ and $b \leftrightarrow c$.

\section{The resummed gluon propagator in a constant background field}\la{resum}

Given the above result for the gluon self-energies, the resummed gluon propagator ${\tilde{D}}_{\mu\nu}^{ab,cd}(P^{ab})$ can be determined with the Dyson-Schwinger equation. In the following, we use the gluon self-energy in the effective theory $\Pi_{{\rm eff};\mu\nu}^{ab,cd}(P^{ab})$ as an example to illustrate the calculation. Technically, there is not anything new in our computation when replacing $\Pi_{{\rm eff};\mu\nu}^{ab,cd}(P^{ab})$ with $\Pi_{{\rm cons};\mu\nu}^{ab,cd}(P^{ab})$ or $\Pi_{{\rm pert};\mu\nu}^{ab,cd}(P^{ab})$.

In covariant gauge, the inverse propagator can be formally written as
\ba
\label{inpro}
&&(\tilde{D}^{-1})_{\mu\nu}^{ab,cd}(P^{ab})=\left[(P^{ab})^2\delta_{\mu\nu}-P_\mu^{ab}P_\nu^{ab}(1-\frac{1}{\xi})\right]{\cal P}^{ab,cd}+\Pi_{\mu\nu}^{ab,cd}(P^{ab})\nonumber\\
&&\equiv {\cal{A}}^{ab,cd}A_{\mu\nu}(P^{ab})+{\cal{B}}^{ab,cd}B_{\mu\nu}(P^{ab})+{\cal{C}}^{ab,cd}P^{ab}_\mu P^{ab}_\nu\, ,
\ea
where
\ba\la{abc}
{\cal{A}}^{ab,cd}&=&\delta^{ad}\delta^{bc}\bigg[ (P^{ab})^2 + F_T(q^a,q^b,P^{ab})\bigg]+\delta^{ab}\delta^{cd}\bigg[-\frac{1}{N} P^2 + G_T(q^a,q^c,P)\bigg]\,,\nonumber\\
{\cal{B}}^{ab,cd}&=&\delta^{ad}\delta^{bc}\bigg[(p^{ab}_0)^2+F_L(q^a,q^b,P^{ab})\bigg]+\delta^{ab}\delta^{cd}\bigg[-\frac{1}{N} p_0^2 + G_L(q^a,q^c,P)\bigg]\,,\nonumber\\
{\cal{C}}^{ab,cd}&=&\delta^{ad}\delta^{bc}\frac{1}{\xi} -\delta^{ab}\delta^{cd}\frac{1}{N\xi}\,.
\ea
In the above equation, the modified structure functions are defined by
\ba
\label{funs}
F_{T/L}(q^a,q^b,P^{ab}) &=& g^2 \Pi_{T/L}(P^{ab})\Big[T^2 \sum_{e}\left( B_2(q^{ae}) + B_2(q^{eb})\right)+C N/(4\pi^2)\Big]\, ,\nonumber \\
G_{T/L}(q^a,q^c,P)&=&- g^2 \Pi_{T/L}(P)\Big[2 T^2 B_2(q^{ac}) + C/(4\pi^2)\Big]\, .
\ea

Due to the transversality of the gluon self-energy, the Lorentz structure of the resummed propagators is a trivial generalization of that in $A_0^{\rm cl}=0$ where three projection operators $A_{\mu\nu}(P^{ab})$, $B_{\mu\nu}(P^{ab})$ and $P_\mu^{ab} P_{\nu}^{ab}$ are all orthogonal to each other. In addition, due to the symmetries of $\Pi_{\mu\nu}^{ab,cd}(P^{ab})$, the inverse propagator $(\tilde{D}^{-1})_{\mu\nu}^{ab,cd}(P^{ab})$ is also invariant under the following exchanges of indices, $\mu \leftrightarrow\nu$, $a \leftrightarrow d$ and $b \leftrightarrow c$. Using the fact that $\sum_c G_{T/L}(q^a,q^c,P)=-F_{T/L}(q^a,q^a,P)$, we find $\sum_{c}{\cal A}^{ab,cc}=\sum_{c}{\cal A}^{cc,ab}=0$ and the same relation holds for ${\cal{B}}^{ab,cd}$ and ${\cal{C}}^{ab,cd}$. Therefore, one can easily show the following identities
\be\la{reid}
\sum_{c}(\tilde{D}^{-1})_{\mu\nu}^{ab,cc}(P)=\sum_{c}(\tilde{D}^{-1})_{\mu\nu}^{cc,ab}(P)=0\,.
\ee
Eq.~(\ref{reid}) is essential which ensures the resummed gluon propagators share the very similar properties as the bare ones. As a result, the corresponding discussions in Sec.~\ref{ba} can be generalized to the resummed solutions straightforwardly. In general, the resummed gluon propagator can be written as
\be\la{gerepro}
\tilde{D}^{ab,cd}_{\mu\nu}(P^{ab})={\cal {X}}^{ab,cd}A_{\mu\nu}(P^{ab})+{\cal {Y}}^{ab,cd}B_{\mu\nu}(P^{ab})+{\cal {Z}}^{ab,cd}P_\mu^{ab} P_{\nu}^{ab}\, .
\ee
In the rest of this section, the calculations of $\tilde{D}^{ab,cd}_{\mu\nu}(P^{ab})$ for diagonal and off-diagonal gluons will be carried out separately.

\subsection{Resummed propagators for off-diagonal gluons}

For off-diagonal gluons, the color structure in $\tilde{D}^{ab,cd}_{\mu\nu}(P^{ab})$ with $a\neq b$  turns to be simple which is similar as that in Eq.~(\ref{did1}). From the basic definition,
\ba\la{indef}
\sum_{ef,\sigma}(\tilde{D}^{-1})_{\mu\sigma}^{ab,ef}(P^{ab})\tilde{D}_{\sigma\nu}^{fe,cd}(P^{fe})&\xlongequal[]{a\neq b}&\sum_{\sigma}(\tilde{D}^{-1})_{\mu\sigma}^{ab,ba}(P^{ab})
\tilde{D}_{\sigma\nu}^{ab,cd}(P^{ab})\nonumber \\
&=&{\cal P}^{ab,cd}\delta_{\mu\nu}\xlongequal[]{a\neq b}\delta^{ad}\delta^{bc}\delta_{\mu\nu}\, ,
\ea
we can get
\ba\la{49}
&&{\cal {X}}^{ab,cd}\Big[(P^{ab})^2 + F_T(q^a,q^b,P^{ab})\Big]A_{\mu\nu}(P^{ab})+\frac{1}{\xi}{\cal {Z}}^{ab,cd}P^{ab}_\mu P^{ab}_\nu  \nonumber \\
&+&\Big[ (p_0^{ab})^2 + F_L(q^a,q^b,P^{ab})\Big]\frac{(P^{ab})^2}{(p_0^{ab})^2}{\cal {Y}}^{ab,cd}B_{\mu\nu}(P^{ab})
\xlongequal[]{a\neq b}\delta^{ad}\delta^{bc}\delta_{\mu\nu}\, .
\ea
It follows that the coefficient of $\delta_{\mu\nu}$ should equal $\delta^{ad}\delta^{bc}$ while the coefficients of the other tensor structures in Lorentz space, for example, of $M_\mu M_\nu$,$M_\mu P_\nu^{ab}$ and $P_\mu^{ab} P_\nu^{ab}$, should vanish. Therefore, we arrive at the following result
\ba\la{intot}
\tilde{D}_{\mu\nu}^{ab,cd}(P^{ab})&\xlongequal[]{a\neq b}&\delta^{ad}\delta^{bc}\bigg\{\frac{1}{(P^{ab})^2 + F_T(q^a,q^b,P^{ab})}A_{\mu\nu}(P^{ab})\nonumber\\
&+&\frac{(p_0^{ab})^4/(P^{ab})^4}{(p^{ab}_0)^2 + F_L(q^a,q^b,P^{ab})}B_{\mu\nu}(P^{ab})+\frac{\xi}{(P^{ab})^4}P_\mu^{ab}P_\nu^{ab}\bigg\}\, .
\ea

The above expression is symmetric under the exchanges of Lorentz and color indices, $\mu \leftrightarrow\nu$, $a \leftrightarrow d$ and $b \leftrightarrow c$, therefore, the following two matrices are commutable as required, namely
\ba
\sum_{ef,\sigma}(\tilde{D}^{-1})_{\mu\sigma}^{ab,ef}(P^{ab})\tilde{D}_{\sigma\nu}^{fe,cd}(P^{fe})\xlongequal[]{a\neq b}\sum_{ef,\sigma}\tilde{D}_{\mu\sigma}^{ab,ef}(P^{ab})
(\tilde{D}^{-1})_{\sigma\nu}^{fe,cd}(P^{fe})\,.
\ea

In the presence of a non-zero background field, Eq.~(\ref{intot}) can be considered as a natural extension of the resummed propagator in the perturbative QGP\cite{Bellac:2011kqa} because they are structurally similar. As for the corrections from the background field, beside the shift in the energies, {\em i.e.}, $p_0\rightarrow p_0^{ab}$, there are also modifications on the structure functions as given in Eq.~(\ref{funs}).

Obviously, when the holonomous gluon self-energy $\Pi_{{\rm cons};\mu\nu}^{ab,cd}(P^{ab})$ is used, the resummed gluon propagator has the same expression as Eq.~(\ref{intot}) where the modified structure function $F_{T/L}(q^a,q^b,P^{ab})$ is now given by Eq.~(\ref{a2}). On the other hand, due to the lost of the transversality, the corresponding calculation with $\Pi_{{\rm pert};\mu\nu}^{ab,cd}(P^{ab})$ turns to be relatively involved. We present the corresponding results in Appendix \ref{app1}.

\subsection{Resummed propagators for diagonal gluons}\la{red}

The resummed propagators for diagonal gluons behave quite differently from the off-diagonal ones due to the much more complicated color structure. By definition, the diagonal components of the resummed propagator satisfy
\be\la{didef}
\sum_{e,\sigma}(\tilde{D}^{-1})_{\mu\sigma}^{a,e}(P)\tilde{D}_{\sigma\nu}^{e,c}(P)={\cal P}^{a,c}\delta_{\mu\nu}\,.
\ee

Taking into account the relation given in Eq.~(\ref{reid}), the basic method for performing this computation is indeed very similar as what we have done for the bare propagators. For a given $c$, we drop one equation with $a=c$ and the other $N-1$ independent equations correspond to $a=1,\cdots,c-1,c+1,\cdots,N$ can be written as
\be\la{rex}
\sum_{e}{\cal{A}}^{a,e}{\cal {X}}^{e,c}={\cal{A}}^{a,a}({\cal {X}}^{a,c}-{\cal {X}}^{c,c})
+\sum_{e\neq a,c}{\cal{A}}^{a,e}({\cal {X}}^{e,c}-{\cal {X}}^{c,c})=-\frac{1}{N}\, ,
\ee
where the term ${\cal{A}}^{a,c}{\cal {X}}^{c,c}$ has been rewritten as $-\sum_{e\neq c}{\cal{A}}^{a,e}{\cal {X}}^{c,c}$. Unlike the coefficient matrix in Eq.~(\ref{cm}) which is independent on the background field, with the insertion of the gluon self-energy contribution, we can not find a simple expression for the corresponding determinant for general $SU(N)$. In general, the $N^2-N$ unknowns ${\cal {X}}^{a,c}-{\cal {X}}^{c,c}$ in Eq.~(\ref{rex}) are not equal although uniquely determinable by using the Cramer's rule\footnote{Here, we need to make an assumption that the determinant of the coefficient matrix is non-zero.}. Equations for ${\cal {Y}}^{a,c}-{\cal {Y}}^{c,c}$ and ${\cal {Z}}^{a,c}-{\cal {Z}}^{c,c}$ ($c$ is fixed and $a \neq c$) can be obtained in a similar way as
\ba\la{rex2}
{\cal{B}}^{a,a}({\cal{Y}}^{a,c}-{\cal{Y}}^{c,c})
+\sum_{e\neq a,c}{\cal{B}}^{a,e}({\cal{Y}}^{e,c}-{\cal{Y}}^{c,c})&=&-\frac{1}{N}\frac{p_0^4}{P^4}\, , \\
\la{rex3}
(1-\frac{1}{N})({\cal{Z}}^{a,c}-{\cal{Z}}^{c,c})-\sum_{e\neq a,c}\frac{1}{N}({\cal{Z}}^{e,c}-{\cal{Z}}^{c,c})&=&-\frac{\xi}{N}\frac{1}{P^4}\, ,
\ea
Notice that the equations for ${\cal{Z}}^{a,c}-{\cal{Z}}^{c,c}$ have no dependence on the background field which are identical to Eq.~(\ref{eqz}) for ${\cal{Z}}_0^{a,c}-{\cal{Z}}_0^{c,c}$.

To proceed further, we also consider
\be\la{ddinre}
\sum_{e,\sigma}\tilde{D}_{\mu\sigma}^{a,e}(P)({\tilde{D}}^{-1})_{\sigma\nu}^{e,c}(P)={\cal P}^{a,c}\delta_{\mu\nu}\, ,
\ee
which leads to the following equation for ${\cal{X}}^{a,c}-{\cal{X}}^{a,a}$
\be\la{rexin}
{\cal{A}}^{c,c}({\cal{X}}^{a,c}-{\cal{X}}^{a,a})
+\sum_{e\neq a,c}{\cal{A}}^{e,c}({\cal{X}}^{a,e}-{\cal{X}}^{a,a})=-\frac{1}{N}\, .
\ee
Comparing with Eq.~(\ref{rex}) and using the fact that ${\cal{A}}^{a,b}={\cal{A}}^{b,a}$, one can show ${\cal{X}}^{a,b}={\cal{X}}^{b,a}$. Similarly, ${\cal{Y}}^{a,b}$ is also unchanged when we flip the color indices, ${\cal{Y}}^{a,b}={\cal{Y}}^{b,a}$.

According to Eq.~(\ref{ddinre}), ${\cal{Z}}^{a,c}-{\cal{Z}}^{a,a}$ also satisfies the same equations as the bare ones. Together with Eq.~(\ref{rex3}), we arrive at
\be\la{recaly}
{\cal{Z}}^{c,a}={\cal{Z}}^{a,c}={\cal{Z}}^{a,a}-\frac{\xi}{P^4}={\cal{Z}}^{c,c}-\frac{\xi}{P^4}\, ,\quad\quad\quad (a \neq c) \,.
\ee
On the other hand, for general $N$, explicit solutions for Eqs.~(\ref{rex}) and (\ref{rex2}) can not be obtained although they are formally solvable. The reason is that calculating determinants of the $(N-1)\times(N-1)$ matrices turns to be very hard due to the non-trivial dependence on the background field. One thing to note is that Eq.~(\ref{recaly}) indicates ${\cal{Z}}^{a,a}={\cal{Z}}^{c,c}$ for $a\neq c$. However, ${\cal{X}}^{a,a}-{\cal{X}}^{c,c}$ and ${\cal{Y}}^{a,a}-{\cal{Y}}^{c,c}$ in general depend on the background field and vanish only when $q\rightarrow0$.

In order to demonstrate the above method in a more explicit way, we take $SU(3)$ as an example to calculate the resummed propagators for diagonal gluons in appendix \ref{appa}.  Besides ${\tilde{D}}_{\mu\nu}^{a,c}-{\tilde{D}}_{\mu\nu}^{c,c}$, the results for individual components ${\tilde{D}}_{\mu\nu}^{a,c}$, which could be useful in other related studies, are also obtained under an extra constraint $\sum_e {\tilde D}_{\mu\nu}^{e,c}(P)=0$. The generalization to arbitrary $N$ is in principle straightforward, however, as just mentioned, computation of the determinants of large matrices will be the major obstacle.

Multiplied by the projection operator, the color summation $\sum_{abcd}{\cal P}^{ab,cd} {\tilde{D}}_{\mu\nu}^{ab,cd}(P^{ab})$ is a quantity of particular interest which can also be uniquely determined by following a similar discussion as the bare ones,
\ba\la{pd}
\sum_{abcd}{\cal P}^{ab,cd} {\tilde{D}}_{\mu\nu}^{ab,cd}(P^{ab})
&=&{\sum_{ab}}^{\prime}{\tilde{D}}_{\mu\nu}^{ab,ba}(P^{ab})-\frac{1}{N}\bigg[{\sum_{ac}}^{\prime}{\tilde{D}}_{\mu\nu}^{a,c}(P)-(N-1)\sum_{a}{\tilde{D}}_{\mu\nu}^{a,a}(P)\bigg]\nonumber\\
&=&{\sum_{ab}}^{\prime}{\tilde{D}}_{\mu\nu}^{ab,ba}(P^{ab})-\frac{1}{N}\bigg[2{\sum_{a>c}}{\tilde{D}}_{\mu\nu}^{a,c}(P)-(N-1)\sum_{a}{\tilde{D}}_{\mu\nu}^{a,a}(P)\bigg]\nonumber\\
&=&{\sum_{ab}}^{\prime}{\tilde{D}}_{\mu\nu}^{ab,ba}(P^{ab})-\frac{1}{N}{\sum_{a>c}}\bigg[{\tilde{D}}_{\mu\nu}^{a,c}(P)+{\tilde{D}}_{\mu\nu}^{c,a}(P)-{\tilde{D}}_{\mu\nu}^{a,a}(P)
-{\tilde{D}}_{\mu\nu}^{c,c}(P)\bigg]\, .\nonumber\\
\ea
In the above equation, terms containing off-diagonal components can be determined by Eq.~(\ref{intot}). In addition, contributions from other terms with diagonal components can be expressed as
\ba\la{udd}
\sum_{a,b}{\cal P}^{a,b} {\tilde{D}}_{\mu\nu}^{a,b}(P)&=&-\frac{1}{N}{\sum_{a>b}}\bigg[{\tilde{D}}_{\mu\nu}^{a,b}(P)+{\tilde{D}}_{\mu\nu}^{b,a}(P)-{\tilde{D}}_{\mu\nu}^{a,a}(P)
-{\tilde{D}}_{\mu\nu}^{b,b}(P)\bigg]\nonumber\\
&=&(N-1)\frac{\xi}{P^4}P_\mu P_\nu
+\frac{1}{{\tilde P}^{2}}\frac{\sum_{k=0}^{N-2} \big(\frac{6 }{N} \big)^k \big(\frac{1 }{1+\beta} \big)^k\Big(1-\frac{P^2}{{\tilde P}^{2}}\Big)^k(N-k-1){\tilde S}_k}{\sum_{k=0}^{N-1} \big(\frac{6 }{N} \big)^k \big(\frac{1 }{1+\beta} \big)^k \Big(1-\frac{P^2}{{\tilde P}^{2}}\Big)^k{\tilde S}_k}A_{\mu\nu}(P)\nonumber\\
&+&\frac{(p_0)^4/P^4}{{\tilde p}_0^{2}}\frac{\sum_{k=0}^{N-2}\big(\frac{6 }{N}  \big)^k \big(\frac{1 }{1+\beta} \big)^k \Big(1-\frac{p_0^2}{{\tilde p}_0^{2}}\Big)^k(N-k-1){\tilde S}_k}{\sum_{k=0}^{N-1}\big(\frac{6 }{N}\big)^k \big(\frac{1 }{1+\beta} \big)^k\Big(1-\frac{p_0^2}{{\tilde p}_0^{2}}\Big)^k{\tilde S}_k}B_{\mu\nu}(P)\, ,
\ea
where
\be\la{newp}
{\tilde P}^2=P^2+\frac{N}{3} g^2 {\tilde T}^2 \Pi_T(P)\, \quad\quad {\rm and} \quad\quad {{\tilde p}_0}^2=p_0^2+\frac{N}{3} g^2 {\tilde T}^2 \Pi_L(P)\, ,
\ee
with ${\tilde{T}}^2 = T^2(1+\beta)$ and $\beta\equiv 3 C/(4\pi^2 T^2)$. In the above equation, the summation of a series of determinants is defined as
\ba\la{det}
\la{tildeS}
{\tilde S}_k = {\sum_{a_{\{k\}}}}
 \begin{vmatrix}
{\tilde{\cal {A}}}^{a_1,a_1} & {\tilde{\cal {A}}}^{a_1,a_2}  & \cdots   & {\tilde{\cal {A}}}^{a_1,a_k}   \\
{\tilde{\cal {A}}}^{a_2,a_1} & {\tilde{\cal {A}}}^{a_2,a_2}  & \cdots   & {\tilde{\cal {A}}}^{a_2,a_k}  \\
\vdots & \vdots  & \ddots   & \vdots  \\
{\tilde{\cal {A}}}^{a_k,a_1} & {\tilde{\cal {A}}}^{a_k,a_2}  & \cdots  & {\tilde{\cal {A}}}^{a_k,a_k}
\end{vmatrix}\, ,
\ea
with
\be
\la{tildeA}{\tilde{\cal {A}}}^{a,b}
= \sum_{e} {\hat B}_2(q^{a e})\delta^{ab}-{\hat B}_2(q^{a b})(1-\delta^{ab})\,, \quad \quad
{\hat B}_2(q^{a b})=B_2(q^{a b})-\frac{1}{6}\,.
\ee
In addition, the shorthand notation ${a_{\{k\}}}$ denotes a set of indices $a_1,a_2,\cdots\,a_k$ which run from $1$ to $N$ and satisfy $a_1<a_2 <\cdots\,<a_k$. Details about the derivation of Eq.~(\ref{udd}) can be found in Appendix \ref{appb}.

We emphasize that Eq.~(\ref{udd}) only presents a formal solution for the diagonal contributions which have rather complicated dependences on the background field. However, the virtue of Eq.~(\ref{udd}) lies in the fact that all the $q$-dependences have been cast into the determinants which vanish in the limit $q\rightarrow 0$. For a given $k$, there are ${\rm C}^{\,k}_N$ (the binomial coefficient) terms in the summation ${\tilde S}_k$ and complications will dramatically increase when $N$ is getting larger and larger.

An alternative way to present the background field effect is to introduce a new set of variables $\lambda_i$ with $i=1,2,\cdots,N-1$. These variables have a nontrivial dependence on the determinant summations ${\tilde S}_k$ which is given by the following equations,
\begin{equation}\label{lambda}
{\sum_{a_{\{k\}}}}\lambda_{a_1}\lambda_{a_2} \cdots\lambda_{a_k}= \tilde{S}_{k} \, ,\quad\quad k=1,2,\cdots, N-1\, ,
\end{equation}
where ${a_{\{k\}}}$ follows the same definition as before except that the indices $a_1,a_2,\cdots\,a_k$ run from $1$ to $N-1$ in Eq.~(\ref{lambda}). Recall that the elements of the real symmetric matrices involved in $\tilde{S}_{k}$ satisfy $\sum_e{\tilde{\cal {A}}}^{e,c}=0$, it can be proved that $\tilde{S}_{k}$ is positive when $k$ is even and it becomes negative for odd $k$. Consequently, one can straightforwardly show that these new variables $\lambda_i<0$ for $i=1,2,\cdots, N-1$.

In terms of  $\lambda_i$, a more compact form of  $\sum_{a,b}{\cal P}^{a,b} {\tilde{D}}_{\mu\nu}^{a,b}(P)$ can be found as
\ba\la{uddnew}
\sum_{a,b}{\cal P}^{a,b}{\tilde D}_{\mu\nu}^{a,b}(P)&=&\sum_{i=1}^{N-1}\Bigg[\frac{1}{P^2+\frac{N}{3}g^2 {\tilde T}^2 \Pi_T(P)(1+\frac{6}{N (1+\beta)} \lambda_i)}A_{\mu\nu}(P)\\ \nonumber
&+& \frac{(p_0)^4/P^4}{p_0^2+\frac{N}{3}g^2{\tilde T}^2 \Pi_L(P)(1+\frac{6}{N(1+\beta)} \lambda_i)} B_{\mu\nu}(P)+\frac{\xi}{P^4}P_\mu P_\nu\Bigg]\, .
\ea
Such an expression is certainly more significant because it can be considered as an analogue to the resummed propagator in the perturbative QGP with vanishing holonomy where $\lambda_i=0$. Eq.~(\ref{uddnew}) also indicates that non-zero background field modifies the transverse gluon self-energy $\Pi_T(P)$ as well as the longitudinal part $\Pi_L(P)$ in the same manner.

In principle, $\sum_{a,b}{\cal P}^{a,b} {\tilde{D}}_{\mu\nu}^{a,b}(P)$ depends on all the diagonal propagators ${\tilde{D}}_{\mu\nu}^{a,b}(P)$. However, it is interesting to point out that this color summation can be simply expressed in terms of ${\tilde{D}}_{\mu\nu}^{a,a}(P)$ if the special constraint $\sum_e {\tilde D}_{\mu\nu}^{e,a}(P)=0$ with $a=1,2,\cdots, N$ is adopted. For general $SU(N)$, it can be shown that
\be
\sum_{a,b}{\cal P}^{a,b}\tilde{D}^{a,b}_{\mu\nu}(P)=(-\frac{1}{N}){\sum_{a,b}}^\prime\tilde{D}^{a,b}_{\mu\nu}(P)+(1-\frac{1}{N})\sum_{a}\tilde{D}^{a,a}_{\mu\nu}(P)=\sum_{a}\tilde{D}^{a,a}_{\mu\nu}(P)\, .
\ee

For the diagonal components, the non-transverse term $\sim M_{\mu}M_{\nu}$ in $\Pi_{{\rm pert};\mu\nu}^{ab,cd}(P^{ab})$ vanishes because $B_3(x)$ is odd. Therefore, the resummed gluon propagator $\tilde{D}^{a,b}_{\mu\nu}(P)$ obtained from $\Pi_{{\rm pert};\mu\nu}^{ab,cd}(P^{ab})$ is the same as that from $\Pi_{{\rm cons};\mu\nu}^{ab,cd}(P^{ab})$. The corresponding calculation can be carried out in exactly the same way as above and the explicit result is identical to Eq.~({\ref{udd}}) or ({\ref{uddnew}}) where one only needs to set the cut-off scale $C$ to be zero, namely, $\beta=0$ and ${\tilde T}^2=T^2$.

\section{The screening effect in a holonomous plasma}\la{screening}

As a direct application of the obtained results, the resummed gluon propagator obtained in imaginary time can be analytically continued to Minkowski time with $i p_0^{ab} \rightarrow \omega$. We are interested in the static limit $\omega \rightarrow 0$ which provides information about the screening effect in a holonomous plasma. Similar problem for the $SU(2)$ gauge theory has been discussed in Ref.~\cite{Reinosa:2016iml} where special attention was paid on the non-perturbative infrared dynamics which is parameterized by a gluon mass originated from the Gribov ambiguity\cite{Reinosa:2014ooa}. In this work, emphasis is placed on the modifications resulting from a non-zero holonomy on the in-medium screening effect. Therefore, the following discussions are based on the effective theory with two dimensional ghosts where a non-zero holonomy can be generated dynamically through the equation of motion.

The definition of the real-time heavy-quark(HQ) potential through the Fourier transform of ${\tilde D}_{00}^{ab,cd}(\omega\rightarrow0)$  can be formulated as the following\cite{Laine:2006ns}
\ba\la{potdef}
V_{QQ}(r)&=&\frac{(ig\gamma_0)^2}{N}\sum_{{\rm colors}}\int \frac{ d^3{\bf p}}{(2\pi)^3}e^{i {\bf p} \cdot {\bf r}}(t^{ab})_{ef}{\tilde D}_{00}^{ab,cd}(\omega\rightarrow0)(t^{cd})_{fe}\, \nonumber \\
&=&-\frac{g^2}{2N}\sum_{{\rm colors}} \int \frac{ d^3{\bf p}}{(2\pi)^3}e^{i {\bf p} \cdot {\bf r}}{\cal P}^{ab,cd}{\tilde D}_{00}^{ab,cd}(\omega\rightarrow0)\,,
\ea
where the quark-gluon vertex reads $i g \gamma_\mu (t^{cd})_{ab}$ in the double line basis and ${\rm Tr}\,(t^{ab}t^{cd})=\frac{1}{2}{\cal P}^{ab,cd}$. In addition, all the color indices should be summed and an extra $N$ in the denominator denotes the color average of the heavy quarks. Eq.~(\ref{potdef}) also indicates that $\sum_{abcd}{\cal P}^{ab,cd}{\tilde D}_{00}^{ab,cd}(P)$ is actually the quantity in question which, as already shown, can be uniquely determined without any extra constraint.

As a simple example, we first look at the static limit of the bare gluon propagator which is given by $(D_0)_{00}^{ab,cd}(\omega\rightarrow0)={\cal P}^{ab,cd}/p^2$. Notice that $(D_0)_{00}^{ab,cd}(\omega, {\bf p})$ has no $q$-dependence after analytically continued to real time. Using the identity $\sum_{abcd}{\cal P}^{ab,cd}{\cal P}^{ab,cd}=N^2-1$, Eq.~(\ref{potdef}) leads to a Coulomb potential $-\alpha_s C_F /r$ with $\alpha_s=g^2/(4\pi) $ and $C_F=(N^2-1)/(2 N)$.

Another example is to consider the static limit of the resummed gluon propagator ${\tilde D}_{00}^{ab,cd}(\omega\rightarrow0)$ at extremely high temperature where the background field $q\rightarrow 0$. Using Eq.~(\ref{intot}), it can be shown that the $N^2-N$ off-diagonal gluons contribute equally and we get the following familiar form
\be\la{od0}
{\sum_{ab}}^{\prime}{\tilde{D}}_{\mu\nu}^{ab,ba}(P,q\rightarrow0)=(N^2-N)\bigg\{\frac{1}{P^2+\frac{N}{3} g^2 T^2 \Pi_T(P)}A_{\mu\nu}(P)+\frac{(p_0)^4/P^4}{p_0^2+\frac{N}{3} g^2 T^2 \Pi_L(P)}B_{\mu\nu}(P)+\frac{\xi}{P^4}P_\mu P_\nu\bigg\}\, .
\ee
For the diagonal gluons, since both $\lambda_i$ vanish as $q\rightarrow 0$, Eq.~(\ref{uddnew}) reduces to a very simple form as
\be\la{d0}
\sum_{a,b}{\cal P}^{a,b} {\tilde{D}}_{\mu\nu}^{a,b}(P,q\rightarrow 0)=(N-1)\bigg\{\frac{1}{P^2+\frac{N}{3} g^2 T^2 \Pi_T(P)}A_{\mu\nu}(P)+\frac{(p_0)^4/P^4}{p_0^2+\frac{N}{3} g^2 T^2 \Pi_L(P)}B_{\mu\nu}(P)+\frac{\xi}{P^4}P_\mu P_\nu\bigg\}\, .
\ee
In principle, the contribution from two dimensional ghosts in the effective theory can be neglected in the high temperature limit $T \gg \sqrt{C}$, where the variable $\beta=3 C/(4\pi^2 T^2)$ is negligible. Therefore, ${\tilde T}^2$ in Eqs.~(\ref{od0}) and (\ref{d0}) has been replaced by $T^2$ for consistency.

As expected, the result of $\sum_{abcd}{\cal P}^{ab,cd} {\tilde{D}}_{\mu\nu}^{ab,cd}(P,q\rightarrow 0)$ computed in the double line basis is identical to that obtained by using the standard generators of $SU(N)$ where the $N^2-1$ gluons give equal contributions. Analytically continuing to Minkowski time with $i p_0 \rightarrow \omega$ and taking the limit $\omega \rightarrow 0$, we find that
\be
\sum_{abcd}{\cal P}^{ab,cd} {\tilde{D}}_{00}^{ab,cd}(\omega\rightarrow0,q\rightarrow 0)=(N^2-1)\frac{1}{p^2+m_D^2}\, .
\ee
Therefore, the Fourier transform in Eq.~(\ref{potdef}) gives the well-known Debye screened potential $-\alpha_s C_F e^{-r m_D}/r$ in the perturbative QGP.

It turns out that the interaction between the heavy quark and anti-quark is not affected by the presence of the holonomy at tree level. On the other hand, the resummed gluon propagator has a non-trivial dependence on the background field, accordingly, one can expect that the Debye screening in the perturbative QGP would be modified when the non-zero holonomy is taken into account.

\subsection {Screening effect from diagonal gluons}

According to Eq.~(\ref{uddnew}), the static limit of $\sum_{a,b}{\cal P}^{a,b}{\tilde D}_{00}^{a,b}$ can be expressed as
\be\la{uddstanew2}
\sum_{a,b}{\cal P}^{a,b}{\tilde D}_{00}^{a,b}(\omega\rightarrow0)=\sum_{i=1}^{N-1}\frac{1}{p^2+\big({\cal M}^{(i)}_{D}\big)^2}\equiv\sum_{i=1}^{N-1}\frac{1}{p^2+{\tilde m}_D^2(1+\frac{6}{N(1+\beta)}  \lambda_i)}\, ,
\ee
with ${\tilde m}_D^2\equiv N g^2 {\tilde T}^2/3$. As we can see, contributions from each diagonal gluons are inversely proportional to $p^2$ plus a $q$-modified mass square, therefore, the $N-1$ diagonal gluons become distinguishable by the associated screening masses ${\cal M}^{(i)}_{D}$ with $i=1,2,\cdots, N-1$. We start by considering $SU(2)$ gauge theory and parameterizing the diagonal color matrix $A_0^{\rm cl}$ as $q^1=-q^2=s/4$. Due to the periodicity of Bernoulli polynomials, ${\tilde S}_k$ actually depends on a set of variables $\Delta q^{ij}$ defined as $\Delta q^{ij}\equiv |q^{ij}-n^{ij}|$ where $n^{ij}$ is an integer that is closest to the value of $q^{ij}$. Without losing any generality, one can assume $0 \le s \le 1$ and it is easy to compute ${\tilde S}_1$ which can be expressed as ${\tilde S}_1= s^2/2- s$. As a result, the explicit form of Eq.~(\ref{uddstanew2}) for $SU(2)$ reads
\be\la{su2re}
\sum_{a,b}{\cal P}^{a,b}{\tilde D}_{00}^{a,b}(\omega\rightarrow0)=\frac{1}{p^2+\big({\cal M}^{(1)}_{D}\big)^2}\, ,
\ee
where the modified screening mass ${\cal M}^{(1)}_{D}$ is given by
\be\la{remd}
{\cal M}^{(1)}_{D}=m_D\sqrt{1+\beta + 3 s^2/2- 3 s}\, .
\ee
Non-zero holonomy results in corrections on the Debye screening effect in the thermal medium which can be described by the ratio between the modified screening mass ${\cal M}_D$ in a holonomous plasma and the Debye screening mass $m_D$ in the perturbative QGP. It is clear from the above equation that this ratio depends on both the background field $s$ and the parameter $\beta=3 C/(4\pi^2 T^2)$. 
However, these two variables $s$ and $\beta$ are not independent, their relation can be obtained from the equations of motion in the effective theory. For general $SU(N)$, the total effective potential in the holonomous plasma is given by\cite{Hidaka:2020aa}
\be\la{ep}
{\cal V}(q)=\sum_{a,b=1}^N {\cal P}_{ab}^{ab} \Big(\frac{2\pi^2T^4}{3} B_4(|q^{ab}|)+\frac{C T^2}{2} B_2(|q^{ab}|)\Big)\, ,
\ee
which leads to the following equations of motion
\be\la{eom}
\sum_{b=1}^N {\rm sign}(q^{ab})\Big(\frac{8\pi^2T^2}{3} B_3(|q^{ab}|)+C B_1(|q^{ab}|)\Big)=0\, .
\ee
Notice that we will simply adopt a constant parameter $C$ in the following discussions. For quantitatively more reliable results, a refined confining potential should be employed where the parameter $C$ could become $T$-dependent. The same purpose can be achieved by including the effects of wave function renormalizations in the gluons and ghost propagators\cite{Lo:2021qkw,Fukushima:2012qa,Aouane:2011fv}.

For $SU(2)$, the background field $s$ in the deconfined phase is given by $s=(1-\sqrt{1-2\beta})$. While at low temperatures, $s=1$ corresponds to the confining vacuum. By requiring that the phase transition occurs at $T_d$ when ${\cal V}(s=1)={\cal V}(s=(1-\sqrt{1-2\beta(T=T_d)}))$, we can determine the cut-off $C=2\pi^2 T_d^2/3$ which indicates $\beta \le 1/2$ for $T\ge T_d$. Therefore, the ratio ${\cal M}_D/m_D$ takes the following simple form
\be\la{rasu2}
{\cal M}^{(1)}_D/m_D=\sqrt{1-T_d^2/T^2}\, .
\ee
Performing the Fourier transform, Eq.~(\ref{su2re}) also leads to a Debye screened potential in the deconfined phase. According to Eq.~(\ref{rasu2}), the screening effect is reduced in a holonomous plasma. In the high temperature limit, the non-perturbative contribution $\sim C$ in the effective theory can be neglected and ${\cal M}^{(1)}_D$ becomes identical to the perturbative $m_D$. In addition, when the temperature approaches to $T_d$ from above, {\em i.e.}, $T\rightarrow T_d^+$, the modified screening mass ${\cal M}^{(1)}_D$ drops to zero smoothly and a vacuum Coulomb potential arises at the deconfinement temperature. Because the background field $s=1$ in the confining vacuum, when $T$ approaches to $T_d$ from below, {\em i.e.}, $T\rightarrow T_d^-$, we also find a vanishing screening mass for the diagonal gluon according to Eq.~(\ref{remd}) where $\beta =1/2$. Therefore, ${\cal M}^{(1)}_D$ is continuous at the critical point, in accord with the fact that the phase transition is second order for $SU(2)$.

Next, we consider the screening effect for $SU(3)$ where we have more than one diagonal gluon. In general, we parameterize the diagonal color matrix $A_0^{\rm cl}$ as $q^1=-q^3=s/3$ and $q^2=0$, which actually corresponds to a real-valued Polyakov loop. Similarly as before, one assumes $0 \le s \le 1$ and the explicit results of ${\tilde S}_1$ and ${\tilde S}_2$ are given by
\be
{\tilde S}_1=\frac{4}{3}s (s-2) \, ,\quad\quad {\tilde S}_2= \frac{1}{9}s^2 (3s^2-14 s+15)\, .
\ee

Solving Eq.~(\ref{lambda}) for $\lambda_1$ and $\lambda_2$, we find $\lambda_1=s^2/3-s$ and $\lambda_2=s^2-5s/3$. Therefore, Eq.~(\ref{uddstanew2}) can be written as
\be\la{su3re}
\sum_{a,b}{\cal P}^{a,b}{\tilde D}_{00}^{a,b}(\omega\rightarrow0)=\frac{1}{p^2+\big({\cal M}^{(1)}_{D}\big)^2}+\frac{1}{p^2+\big({\cal M}^{(2)}_{D}\big)^2}\,,
\ee
with
\be\la{su3ma}
{\cal M}^{(1)}_{D}= m_D\sqrt{1+\beta+ 2 s^2/3-2s}\quad\quad{\rm and}\quad\quad {\cal M}^{(2)}_{D}= m_D \sqrt{1+\beta+ 2 s^2-10s/3}\,.
\ee
As we can see, the two diagonal gluons can be distinguished by their screening masses in a holonomous plasma. According to the equations of motion,  the background field in the deconfined phase is given by $s=(3-\sqrt{9-24\beta})/4$. Furthermore, that the first order phase transition happens at the deconfinement temperature determines the value of the cut-off $C=40\pi^2 T_d^2/81$ which indicates $\beta$ equals $10/27$ when $T=T_d$. Then, it is straightforward to show the following ratio between the modified screening mass and the perturbative $m_D$,
\be\la{rasu3}
{\cal M}^{(1)}_{D}/m_D=\frac{1}{2}\sqrt{1+\sqrt{9-80T_d^2/(9T^2)}}\quad\quad{\rm and}\quad\quad{\cal M}^{(2)}_{D}/m_D=\frac{\sqrt{3}}{6}\sqrt{9-80T_d^2/(9T^2)+\sqrt{9-80T_d^2/(9T^2)}}\,.
\ee

In the deconfined phase, ${\cal M}^{(1)}_{D}$ is always larger than ${\cal M}^{(2)}_{D}$ and they become identical to $m_D$ only in the limit $T\rightarrow \infty$. Eq.~(\ref{rasu3}) also shows that both of the modified screening masses are smaller than the perturbative Debye mass $m_D$, thus a reduced screening effect can be expected in a holonomous plasma. However, at the deconfinement temperature, neither of modified screening masses vanishes which is different from the behavior found in $SU(2)$. In fact, we find ${\cal M}^{(1)}_{D}/m_D=\sqrt{3}/3$ and ${\cal M}^{(2)}_{D}/m_D=\sqrt{3}/9$ when $T\rightarrow T_d^+$. Therefore, a Debye screened potential persists in the deconfined phase and no vacuum Coulomb potential shows up at the deconfinement temperature. On the other hand, in the confined phase where $s=1$, the two modified screening masses as given in Eq.~(\ref{su3ma}) are the same and ${\cal M}^{(1)}_{D}/m_D={\cal M}^{(2)}_{D}/m_D=\sqrt{3}/9$ when $T\rightarrow T_d^-$\footnote{However, this is not true for general $SU(N)$. For example, under the straight line {\em ansatz} Eq.~(\ref{sla}), we find that only two of three screening masses (for diagonal gluons) in $SU(4)$ become identical when $T\rightarrow T_d^-$.}. The jump in the screening mass ${\cal M}^{(1)}_{D}$ at the critical point reflects the nature of the first order phase transition in $SU(3)$ gauge theory. This is significantly different from the second-order phase transition in $SU(2)$ where the modified screening mass becomes continuous. The same behavior is also found in the lattice simulations\cite{Maas:2011ez}.

To generalize the above results to $SU(N)$, we need to calculate the determinants in $\tilde{S}_{k}$ and solve the equations for $\lambda_i$ as given in Eq.~(\ref{lambda}) for arbitrary $N$. However, this would be rather tedious when $N$ is large. In addition, the background field in general can not be parameterized with a signal variable for $N>3$ which further complicates the situation. To proceed further, we will focus on the high temperature region where $\Delta q^{ij}\ll 1$, then the dominant contribution from ${\tilde S}_k$ is proportional to the $k^{\rm th}$ power of $\Delta q^{ij}$ and gets suppressed when $k$ is large. Within the leading order approximation, only ${\tilde S}_1$ contributes while ${\tilde S}_k$ with $k>1$ becomes negligible. Thus Eq.~(\ref{lambda}) is simplified into $\lambda_1+\lambda_2+\cdots+\lambda_{N-1}={\tilde S}_1\approx - 2\sum_{i<j} \Delta q^{ij}$. Since there is no unique solution for $\lambda_i$ in this case, it is natural to assume that all the $N-1$ $\lambda_i$'s are equal. Under this assumption, however, the $N-1$ diagonal gluons are no longer distinguishable by their modified screening masses. Consequently, Eq.~(\ref{uddstanew2}) takes the following form
\be\la{debye2}
\sum_{a,b}{\cal P}^{a,b}{\tilde D}_{00}^{a,b}(\omega\rightarrow0)\approx (N-1)\frac{1}{p^2+{\cal M}_D^2}\,,
\ee
where the modified screening mass ${\cal M}_D^2$ is given by
\be\la{remdnew}
{\cal M}_D^2=m_D^2\Big[1+\beta-\frac{12}{N(N-1)}\sum_{i<j} \Delta q^{ij}\Big]\, .
\ee

As mentioned before, the parameter $\beta$ is related to the background field $q$ via the equations of motion. For general $SU(N)$, we adopt the straight line {\em ansatz} for the background field\cite{Dumitru:2012fw}
\be\la{sla}
q^i=\frac{N-2i+1}{2N} s\, ,
\ee
which satisfies the constraint $\sum_{i=1}^{N}q^i=0$ and also leads to a real valued Polyakov loop. In the above equation, $0\le s \le 1$ and the perturbative vacuum corresponds to $s=0$ while the confining vacuum is at $s=1$. Notice that Eq.~(\ref{sla}) corresponds to the exact solutions for two and three colors. For $N>3$, the deviation from the straight line turns to be very small\cite{Dumitru:2012fw}. Since our discussion here applies at high temperature, therefore, we consider $s \ll 1$ in order to be consistent with the previous assumption $\Delta q^{ij}\ll 1$. Solving Eq.~(\ref{eom}) with the above {\em ansatz}, the following identity can be derived
\be
s=1-\frac{1}{8(1-3/(2N^2))}\Big(3(1-\frac{4}{N^2})+\sqrt{25-80(1-\frac{3}{2N^2})\beta}\Big)\, .
\ee
Equivalently, we find $\beta\approx s$ up to linear order in $s$. According to the above discussions, the modified screening mass as given in Eq.~(\ref{remdnew}) can be expressed as
\be\la{remd2}
{\cal M}_D^2=m_D^2\big(1-\frac{N+2}{N} s\big)\, ,\quad\quad {\rm for}\quad\quad s\ll1\, ,
\ee
where we have used $\sum_{i<j} \Delta q^{ij}=(N^2-1)s/6$. As we can see, the modified screening mass is reduced for non-zero $s$ and the deviation from $m_D$ becomes smaller when $N$ increases. Comparing Eq.~(\ref{remd2}) with Eq.~(\ref{remd}) in the limit $s\ll 1$, it is direct to see the equivalence for $SU(2)$. For $SU(3)$, Eq.~(\ref{su3ma}) leads to two different screening masses ${\cal M}_{1D}/m_D=\sqrt{1-s}$ and ${\cal M}_{2D}/m_D=\sqrt{1-7s/3}$ for $s\ll 1$. On the other hand, Eq.~(\ref{remd2}) shows a modified screening mass ${\cal M}_D/m_D=\sqrt{1-5s/3}$. In fact, the equivalence can be shown by looking at the corrections to the Debye screening potential in the perturbative QGP, as in both cases, the corrections up to linear order in $s$ are the same and equal to $- 5 s \alpha_s m_D {\rm e}  ^{-r m_D} / 18$.

Based on the above analysis, the following conclusion can be drawn for general $SU(N)$ that is introducing a small but non-zero background field merely amounts to modifications on the perturbative Debye mass $m_D$ and the corresponding HQ potential is always deeper than the perturbative screened potential characterized by $m_D$, which suggests a weaker screening, thus a more tightly bounded quarkonium state in a holonomous plasma. In addition, performing the Fourier transform, the resulting HQ potential remains the standard Debye screened form which can be expressed as\footnote{In this subsection, $V_{QQ}(r)$ actually refers to the HQ potential associated with diagonal gluons, namely, terms with $a\neq b$ are excluded  in the color summation in Eq.~(\ref{potdef}).}
\be
V_{QQ}(r, s\ll 1)=-\frac{N-1}{2N}\frac{\alpha_s}{r} e ^{-r m_D \sqrt{1-\frac{N+2}{N} s}}\, .
\ee

The above discussions for the diagonal gluons are based on the effective theory for a holonomous plasma with contributions from two dimensional ghosts. When the resummed gluon propagators $\tilde{D}^{a,b}_{\mu\nu}(P)$ obtained from $\Pi_{{\rm pert};\mu\nu}^{ab,cd}(P^{ab})$ or $\Pi_{{\rm cons};\mu\nu}^{ab,cd}(P^{ab})$ are considered, the corresponding analysis turns to be very similar and one only needs to set $\beta=0$ in Eq.~(\ref{uddstanew2}). However, we would like to point out that the above results depend on the use of the equations of motion which generate non-zero holonomy at any finite temperature in the effective theory. On the other hand, dropping the contributions from two dimensional ghosts, the system would be always in the perturbative vacuum which actually corresponds to vanishing holonomy. In this case, non-zero holonomy has to be introduced by hand which doesn't obey the corresponding equations of motion in the perturbation theory. In particular, the modified screening mass square could be negative with certain values of the background field and this doesn't appear in our above discussions. The necessity of only looking at solutions that satisfy the equations of motion was also found in the related studies\cite{KorthalsAltes:2020ryu,Hidaka:2020aa,KorthalsAltes:2019yih}.

\subsection {Screening effect from off-diagonal gluons}

The resummed propagator for off-diagonal gluons has a relatively simple form as given in Eq.~(\ref{intot}) which doesn't contain complicated determinants. After analytically continuing to Minkowski time, we get the following result for the temporal component of the resummed propagator
\be\la{totch}
{\tilde D}_{00}^{ab,ba}(\omega\rightarrow0)=\frac{1}{p^2+g^2 \Big[T^2 \sum_{e}\left( B_2(q^{ae}) + B_2(q^{eb})\right)+C N/(4\pi^2)\Big]}\, .
\ee
Formally, we can also define the modified screening mass for each off-diagonal gluons
\be\la{moffd}
{({\cal M}_D^{(ab)})}^2=m_D^2\Big[1+\frac{3}{N} \sum_{e}\left( {\hat B}_2(q^{ae}) + {\hat B}_2(q^{eb})\right)+\beta\Big] \quad{\rm for}\quad a\neq b\, ,
\ee
which reduces to the perturbative $m_D^2$ in the high temperature limit where $T\gg \sqrt {C}$ and the background field $q\rightarrow 0$.

For $SU(2)$, the two off-diagonal gluons have the same modified screening mass. Taking $q^1=-q^2=4/s$, it is easy to show
\be
{\cal M}_D^{(12)}={\cal M}_D^{(21)}=m_D\sqrt{1+\beta+3s^2/4-3s/2}\, .
\ee
Following what we have done for the diagonal gluons, one should further take into account the equations of motion $s=(1-\sqrt{1-2\beta})$ and choose $C=2\pi^2 T_d^2/3$, thus the temperature dependence of ${\cal M}_D^{(12)}$ is found to be
\be
{\cal M}_D^{(12)}={\cal M}_D^{(21)}=m_D\sqrt{1-T_d^2/(4T^2)}\, .
\ee
Comparing with Eq.~(\ref{rasu2}), we find that at a given temperature, the screening mass for off-diagonal gluons is larger than the diagonal one, therefore, the former has a smaller reduction in the screening effect. Notice that ${\cal M}_D^{(12)}$ has a non-vanishing value $\sqrt{3}m_D/2$ as $T \rightarrow T_d$ either from above or from below. Therefore, only diagonal gluon is unscreened at the deconfinement temperature for $SU(2)$.

There are six off-diagonal gluons in $SU(3)$, but only two different screening masses which are denoted as ${\cal M}_D^{(23)}$ and ${\cal M}_D^{(13)}$. With the same parameterization of the background field as for the diagonal gluons, we can show that
\be\la{su3off}
{\cal M}_D^{(23)}=m_D\sqrt{1+\beta+7s^2/9-5s/3}\quad\quad{\rm and}\quad\quad{\cal M}_D^{(13)}=m_D\sqrt{1+\beta+10s^2/9-2s}\,.
\ee
Imposing the following conditions $s=(3-\sqrt{9-24\beta})/4$ and $C=40\pi^2 T_d^2/81$, we arrive at
\ba
{\cal M}_D^{(23)}&=&m_D\frac{\sqrt{405-40(T_d/T)^2+27\sqrt{81-80(T_d/T)^2}}}{18\sqrt{2}}\, ,\nonumber \\
{\cal M}_D^{(13)}&=&m_D\frac{\sqrt{243-80(T_d/T)^2+9\sqrt{81-80(T_d/T)^2}}}{18}\,.
\ea
At a given temperature in the deconfined phase, ${\cal M}_D^{(23)}$ is always larger than ${\cal M}_D^{(13)}$ and these two modified screening masses are both smaller than the perturbative $m_D$. Similar as $SU(2)$, off-diagonal gluons show a stronger screening effect as compared to the diagonal ones since their screening masses are larger than those given in Eq.~(\ref{rasu3}). In addition, we find ${\cal M}_D^{(23)}/m_D=7/9$ and ${\cal M}_D^{(13)}/m_D=\sqrt{43}/9$ as $T\rightarrow T_d^+$. While in the confined phase with $s=1$, the two screening masses in Eq.~(\ref{su3off}) have the same value and ${\cal M}_{D}^{(23)}/m_D={\cal M}_{D}^{(13)}/m_D=\sqrt{39}/9$ as $T\rightarrow T_d^-$. It is clear that  there is also a jump in the screening masses for off-diagonal gluons at the critical point.

For general $SU(N)$, given the straight line {\em ansatz} Eq.~(\ref{sla}), one can also derive an analytical expression for the screening mass ${\cal M}_D^{(ab)}$ in the deconfined phase which, due to the rather complicated from, is not list here.  However, in the high temperature limit where $s$ is small enough, we can show that
\be
{\cal M}_D^{(ab)}/m_D=\sqrt{1+\frac{s}{N^2}\big[3(N+1)(a+b)-3(a^2+b^2)-(2N^2+3N)\big]}\quad{\rm for}\quad s\ll 1\, .
\ee
In the above equation, the small but non-zero background field $s$  leads to a reduced screening mass ${\cal M}_D^{(ab)}$, therefore, the screening effect related to the off-diagonal gluons is also weakened in a holonomous plasma.

Furthermore, we can study the behavior of ${\cal M}_D^{(ab)}$ at the deconfinement temperature. In the confined phase, the modified screening mass is simplified to ${\cal M}_D^{(ab)}/m_D=\sqrt{1+\beta N^2}/N$. Therefore, there is only one screening mass for all the off-diagonal gluons as $T\rightarrow T_d^-$. Explicitly, we have
\be
{\cal M}_D^{(ab)}=m_D\sqrt{(3N^4+11N^2-17)/(10N^2-15)}/N\quad{\rm for}\quad T\rightarrow T_d^-\,.
\ee
Instead of showing the corresponding result at $T\rightarrow T_d^+$, we can look at the jump in the screening masses at the critical point which exists for $N>2$ and is given by the following expression,
\ba
\Delta\Big(\frac{{\cal M}_D^{(ab)}}{m_D}\Big)^2&\equiv&\left.\Big(\frac{{\cal M}_D^{(ab)}}{m_D}\Big)^2\right |_{T\rightarrow T_d^+}-\left.\Big(\frac{{\cal M}_D^{(ab)}}{m_D}\Big)^2\right |_{T\rightarrow T_d^-}\nonumber \\
&=&\frac{(N^2-4)}{N^2(3-2N^2)^2}\big[3(N^3+N^2+N+1)(a+b)-3(N^2+1)(a^2+b^2)-(3N^3+8N^2+3N-2)\big]\, .
\ea

\begin{figure}[htbp]
\centering
\includegraphics[width=0.45\textwidth]{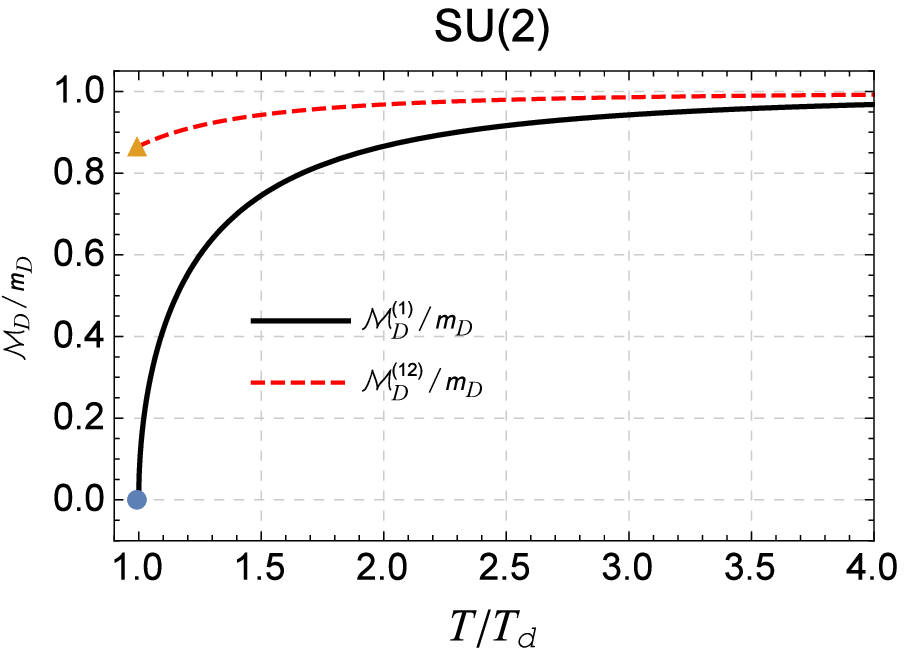}
\includegraphics[width=0.45\textwidth]{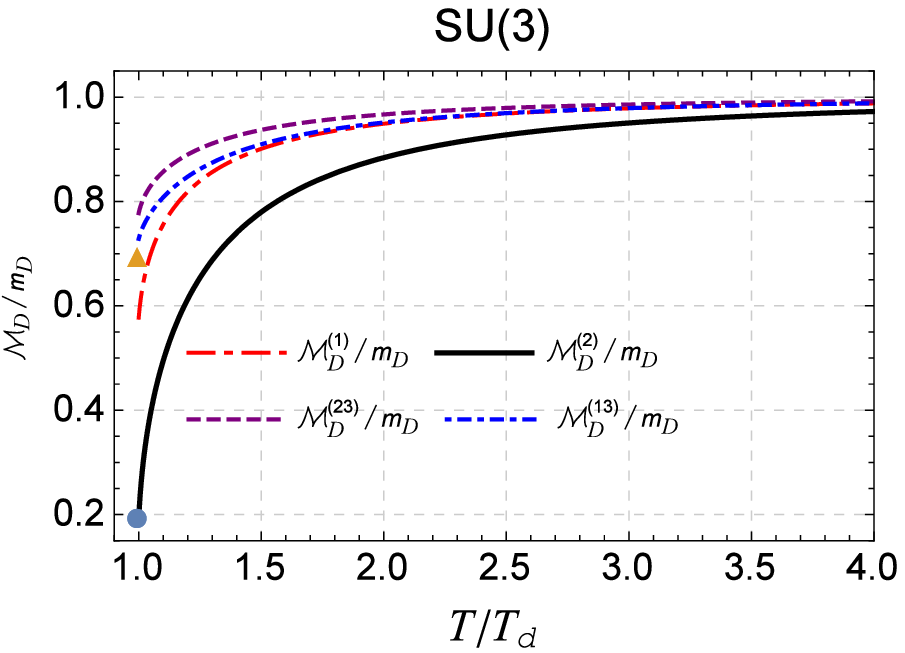}
\vspace*{-.01cm}
\caption{\small {The ratio ${\cal M}_D/m_D$ as a function of $T/T_d$ for $SU(2)$ (left) and $SU(3)$ (right). The corresponding results at $T\rightarrow T_d^-$ are denoted by a circle for diagonal gluons and by a triangle for off-diagonal gluons.}
\label{fig1}}
\end{figure}

\begin{figure}[htbp]
\centering
\includegraphics[width=0.45\textwidth]{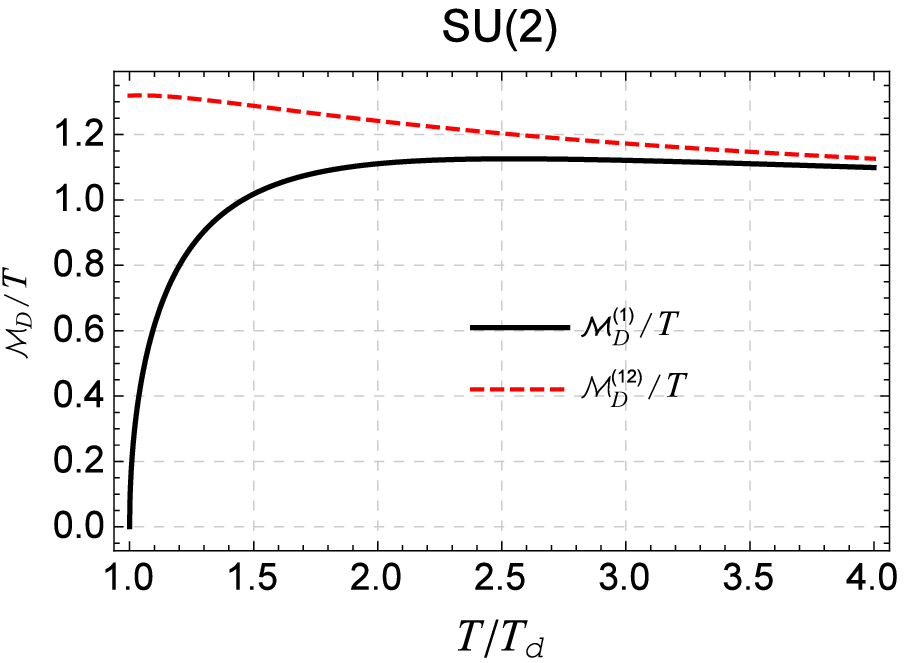}
\includegraphics[width=0.45\textwidth]{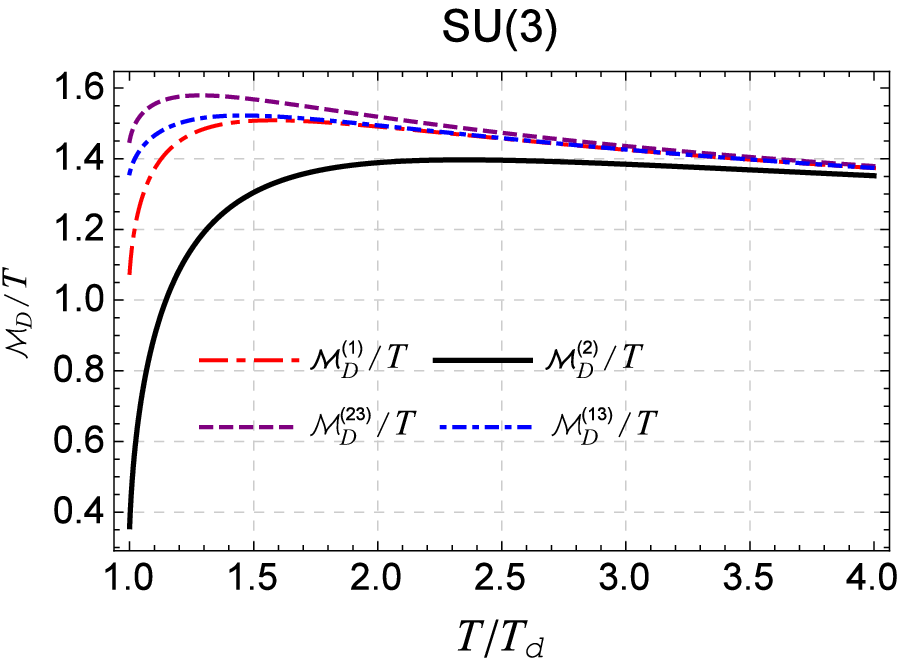}
\vspace*{-.01cm}
\caption{\small {The ratio ${\cal M}_D/T$ as a function of $T/T_d$ for $SU(2)$ (left) and $SU(3)$ (right). For numerical evaluations, we use the 2-loop perturbative running coupling.}
\label{fig2}}
\end{figure}

In Fig.~\ref{fig1}, we show the ratio ${\cal M}_D/m_D$ as a function of $T/T_d$ for $SU(2)$ (left) and $SU(3)$ (right). The corresponding results at $T\rightarrow T_d^-$ are denoted by a circle for diagonal gluons and by a triangle for off-diagonal gluons. Quantitatively, the deviation from unity becomes negligible when $T$ is higher than $\sim 4 T_d$ where the background field is too small to induce visible modifications on the perturbative Debye mass $m_D$. On the contrary, in the ``semi"-QGP region, namely, from $T_d$ to about $4T_d$, a reduced screening is clear to see from these plots. However, it is not possible to make a direct comparison with the lattice simulations where the new feature that the $N^2-1$ gluons are distinguishable by their associated screening masses has not been taken into account. On the other hand, as shown in Fig.~\ref{fig2}, the qualitative behaviors of the ratio ${\cal M}_D/T$ as a function of $T/T_d$ indeed are very similar as those found in the lattice simulations, not only for pure gauge theories\cite{Digal:2003jc} but also for two-flavor QCD\cite{Maezawa:2010vj,Kaczmarek:2005ui}. In general, the ratio ${\cal M}_D/T$ is not a monotonic function of $T$. In the high temperature region where the holonomy is small, it grows with decreasing $T$ which can be understood as a consequence of the increase in the running coupling. There exists a turning point at a temperature close but above $T_d$ where the ratio ${\cal M}_D/T$ starts to fall\footnote{For $SU(2)$, ${\cal M}^{(12)}_D/T$ also decreases with decreasing $T$ when $T$ is very close to $T_d$. This may be not very clear to see from the plot.}. In fact, for temperatures close to $T_d$, the influence of the non-zero holonomy, which leads to the decrease of ${\cal M}_D/T$, becomes dominant over the running effect of the strong coupling which in turn increases the ratio.

It is worth noting that presumably the above discussions on the screening effect in a holonomous plasma are only applicable in the deconfined phase where gluons are the physical degrees of freedom. When applying Eq.~(\ref{ep}) to the confined regime, some of the thermodynamic quantities go negative\cite{Meisinger:2001cq,Dumitru:2012fw}, therefore, a refined effective potential that incorporates contributions from glueballs turns to be important for a consistent analysis at temperatures below $T_d$. Despite the above mentioned issues, a naive generalization of the obtained results to the confined phase can be carried out by  taking the background field $s=1$. Consequently, we find that the ratio ${\cal {M}}_D/m_D$ grows with decreasing $T$ which is in contrast to the observation as shown in Fig.~\ref{fig1}. Furthermore, by assuming the leading order perturbative form $m_D \sim g T$ persists even in the confined regime, the diagonal screening masses ${\cal {M}}_D^{(i)}$ increase with decreasing $T$ provided that the coupling $g$ is fixed. This is qualitatively in agreement with the lattice simulations as shown in Refs.~\cite{Maas:2011ez,Silva:2013maa}. However, the opposite conclusion holds for the off-diagonal screening masses ${\cal {M}}_D^{(ab)}$. In addition, one can show that all the screening masses approach $\sqrt{N \alpha_s C/\pi}$ in the zero-temperature limit. For $SU(3)$, given $T_d=0.27 {\rm GeV}$ and $g\approx 1.87$ (this is the value predicted by the 2-loop running coupling at $T_d$), $\sqrt{N \alpha_s C/\pi} \approx 0.31 {\rm GeV}$ which is comparable to the above mentioned lattice results.

We are not going to discuss the corresponding results based on the holonomous gluon self-energy $\Pi_{{\rm pert};\mu\nu}^{ab,cd}(P^{ab})$ or $\Pi_{{\rm cons};\mu\nu}^{ab,cd}(P^{ab})$ obtained in the perturbation theory. This is because the static limit of the resummed propagator ${\tilde D}_{00}^{ab,ba}(\omega\rightarrow0)$ is not well defined. In fact, it is straightforward to show the following
\ba\la{rept}
{\tilde D}_{{\rm cons};00}^{ab,ba}(\omega\rightarrow0)&=&\frac{1}{p^2+g^2 T^2 \sum_{e}\left( B_2(q^{ae}) + B_2(q^{eb})\right)+\frac{i J^\prime(q^a,q^b)}{2p}\left.\left(\ln\frac{\omega+p}{\omega-p}-\frac{2p}{\omega}\right)\right | _{\omega\rightarrow 0}}  \, , \nonumber \\
{\tilde D}_{{\rm pert};00}^{ab,ba}(\omega\rightarrow0)&=&\frac{1}{p^2+g^2 T^2 \sum_{e}\left( B_2(q^{ae}) + B_2(q^{eb})\right)+\frac{i J^\prime(q^a,q^b)}{2p}\left.\left(\ln\frac{\omega+p}{\omega-p}\right)\right | _{\omega\rightarrow 0} -\frac{\xi}{p^4} \big(J^\prime(q^a,q^b)\big)^2}\, ,
\ea
with
\be
J^\prime(q^a,q^b) =  \frac{4 \pi}{3} g^2T^3 \sum_{e} \left( B_3(q^{ae}) + B_3(q^{eb}) \right)\, .
\ee
With the constraint contribution, the screening mass becomes divergent due to the appearance of an unexpected term $\sim 1/\omega$ in the static limit. On the other hand, using the non-transverse $\Pi_{{\rm pert};\mu\nu}^{ab,cd}(P^{ab})$, the screening mass is gauge-dependent. In addition, the retarded solution $i p_0^{ab}\rightarrow \omega +i \epsilon$ leads to a different result in the static limit as compared to the advanced solution $i p_0^{ab}\rightarrow \omega -i \epsilon$ because for the logarithmic term, we have $\left.\ln\frac{\omega+p\pm i \epsilon}{\omega-p\pm i \epsilon}\right | _{\omega\rightarrow 0}=\mp i \pi$. All of these problems are related to the anomalous term $\sim {\cal K}^{ab, cd}(q)$ or $\sim B_3(x)$ in the gluon self-energy which vanishes when the background field equals zero. This is again an example to show the necessity of only looking at solutions that satisfy the equations of motion.

\section{Summary and Outlook}\la{summary}

In this work, we have computed the resummed gluon propagator in a QCD plasma with non-zero holonomy which was realized by introducing a classical background field for the vector potential $A_0$. Being crucial for many processes with soft momentum exchange, the resummed propagator was obtained through the Dyson-Schwinger equation where, as a necessary input quantity, the gluon self-energy in a holonomous plasma has been calculated previously in an effective theory where non-zero holonomy can be dynamically generated.

Due to the transversality of the gluon self-energy in a constant background field, the resummed propagator for off-diagonal gluons as given in Eq.~(\ref{intot}) is formally analogous to that in the perturbative QGP with vanishing holonomy. The real difficulties in the computation exist in the color structure related to the diagonal gluons. The double line basis, as extensively used before, is convenient to compute in the presence of a background field. However, due to the over-completeness, diagonal gluons are mixed in the double lime basis and the propagator associated with each individual gluon cannot be uniquely determined. Instead, as shown in Eq.~(\ref{uddnew}), the color summation $\sum_{a,b}{\cal P}^{a,b}\tilde{D}^{a,b}_{\mu\nu}(P)$ has a definite expression in which all the background field dependence can be cast into the determinants of a series of matrices and the corresponding evaluation turns to be rather complicated when $N$ is large.

After analytically continued to Minkowski time, the static limit of the resummed gluon propagators was also discussed which offered an insight into the screening effects in a holonomous plasma. In general, introducing non-zero holonomy merely amounts to modifications on the perturbative Debye mass $m_D$ and the resulting HQ potential, which remains the standard Debye screened form, is always deeper than the screened potential in the perturbative QGP. Therefore, a weaker screening, thus a more tightly bounded quarkonium state can be expected in a holonomous plasma. In addition, both the diagonal and off-diagonal gluons become distinguishable by their modified screening masses ${\cal M}_D$ as given in Eqs.~(\ref{uddstanew2}) and (\ref{moffd}), respectively.

The explicit $T$-dependence of the modifications on the perturbative $m_D$ as described by the ratio ${\cal M}_D/m_D$ was derived by imposing the equations of motion for the background field. Taking $SU(2)$ and $SU(3)$ as examples, the deviation of ${\cal M}_D/m_D$ from its high temperature limit where the ratio approaches to one is only dramatic near the deconfinement temperature $T_d$, according to the plots in Fig.~\ref{fig1}. As the temperature decreases to $T_d$, the modified screening masses have non-vanishing values with the only exception of the screening mass ${\cal M}^{(1)}_D$ associated with the diagonal gluon in $SU(2)$, which drops to zero as $T\rightarrow T_d$. Furthermore, there is a jump in the modified screening masses at the deconfinement temperature for $N>2$ and this is naturally expected in first order phase transitions in $SU(N)$ gauge theories. We also discussed the behavior of ${\cal M}_D/T$ as a function of the temperature $T$ which, as shown in Fig.~\ref{fig2}, exhibits a very similar $T$-dependence as observed in lattice simulations. As a non-monotonic function of $T$, the change of ${\cal M}_D/T$ with decreasing temperature can be understood as a competition between the running of the strong coupling which increases ${\cal M}_D/T$ and the influence of the non-zero holonomy, which in turn leads to the decrease of the ratio.

We would like to point out that the above conclusions are based on the use of holonomous gluon self-energy obtained in the effective theory where by embedding two dimensional ghosts isotropically into four dimensions, a new contribution arising in the effective potential ensures a non-zero holonomy at any finite temperature. Dropping such a contribution, the computation of the resummed gluon propagator with holonomous gluon self-energy in perturbation theory doesn't involve anything new as we already discussed. However, the equations of motion suggest vanishing background field in perturbation theory, therefore, deviating from the perturbative vacuum turns to be not self-consistent due to the violation of the equations of motion. 
In particular, even taking $q\rightarrow 0$ may cause problem in the perturbation theory because one would encounter ambiguous expressions of the type ``$0/0$" in the static limit, see Eq.~(\ref{rept}). This indicates that the system has to stay exactly in the perturbative vacuum. As a result, generating non-zero holonomy from the equations of motion is essential in a holonomous plasma, however, perturbation theory fails to do so. Finally, to generalize our computation to full QCD, one should also include a new term in the action analogous to what has been done in the pure gauge theories. It is expected to cancel the same anomalous term $\sim {\cal K}^{ab,cd}(q)$ showing up in the fermionic contributions to the holonomous gluon self-energy. This will be investigated in future work.

\section*{Acknowledgments}
\label{sec:acknowledgments}

We acknowledge helpful discussions with R. D. Pisarski and U. Reinosa. This work is supported by the NSFC of China under Project Nos. 11665008 and 12065004, by Natural Science Foundation of Guangxi Province of China under Project Nos. 2016GXNSFFA380014, 2018GXNSFAA138163 and by the ``Hundred Talents Plan" of Guangxi Province of China.

\appendix

\section{Resummed propagator for off-diagonal gluons obtained from $\Pi_{{\rm pert};\mu\nu}^{ab,cd}(P^{ab})$}
\la{app1}

With the perturbative gluon self-energy $\Pi_{{\rm pert};\mu\nu}^{ab,cd}(P^{ab})$, the inverse propagator in covariant gauge reads
\ba
\la{a1}
&&(\tilde{D}^{-1})_{\mu\nu}^{ab,cd}(P^{ab})=\left[(P^{ab})^2\delta_{\mu\nu}-P_\mu^{ab}P_\nu^{ab}(1-\frac{1}{\xi})\right]{\cal P}^{ab,cd}+\Pi_{{\rm pert};\mu\nu}^{ab,cd}(P^{ab}) \nonumber\\
&&\equiv {\cal{A}}^{ab,cd}A_{\mu\nu}(P^{ab})+{\cal{B}}^{ab,cd}B_{\mu\nu}(P^{ab})+{\cal{C}}^{ab,cd}P^{ab}_\mu P^{ab}_\nu+{\cal{E}}^{ab,cd} M_\mu M_\nu\, ,
\ea
where ${\cal{A}}^{ab,cd}$, ${\cal{B}}^{ab,cd}$ and ${\cal{C}}^{ab,cd}$ are formally the same as those given by Eq.~(\ref{abc}) but the corresponding structure functions now take the following forms
\ba
\label{a2}
F_{T/L}(q^a,q^b,P^{ab})&=& g^2T^2\Pi_{T/L}(P^{ab})\sum_{e}\left( B_2(q^{ae}) + B_2(q^{eb})\right)-\Pi_{T/L}(P^{ab}) J(q^a,q^b,p_0^{ab})\, ,\nonumber \\
G_{T/L}(q^a,q^c,P)&=&-2 g^2 T^2 \Pi_{T/L}(P)  B_2(q^{ac})\, ,
\ea
with
\be
J(q^a,q^b,p_0^{ab}) =  \frac{4 \pi}{3} g^2T^2\frac{T}{p_0^{ab}}\sum_{e} \left( B_3(q^{ae}) + B_3(q^{eb}) \right)\, .
\ee
In addition, the introduced new term in Eq.~(\ref{a1}) is defined as ${\cal{E}}^{ab,cd}=\delta^{ad}\delta^{bc} J(q^a,q^b,p_0^{ab})$.

Assuming the following expression for the resummed gluon propagator
\be\la{a4}
\tilde{D}^{ab,cd}_{\mu\nu}(P^{ab})={\cal {X}}^{ab,cd}A_{\mu\nu}(P^{ab})+{\cal {Y}}^{ab,cd}B_{\mu\nu}(P^{ab})+{\cal {Z}}^{ab,cd}P_\mu^{ab} P_{\nu}^{ab}+{\cal {W}}^{ab,cd}M_\mu M_{\nu}\, ,
\ee
according to Eq.~(\ref{indef}), we arrive at
\ba
&&\big[ (p_0^{ab})^2 + F_L(q^a,q^b,P^{ab})\big]\big[\frac{(P^{ab})^2}{(p_0^{ab})^2}{\cal {Y}}^{ab,cd}B_{\mu\nu}(P^{ab})+{\cal {W}}^{ab,cd} B_{\mu\sigma}(P^{ab})\cdot M_\sigma M_\nu\big]\nonumber \\
&+&{\cal {X}}^{ab,cd}\big[(P^{ab})^2 + F_T(q^a,q^b,P^{ab})\big]A_{\mu\nu}(P^{ab})+\frac{1}{\xi}{\cal {Z}}^{ab,cd}P^{ab}_\mu P^{ab}_\nu+J(q^a,q^b,p_0^{ab}) {\cal {W}}^{ab,cd} M_\mu M_\nu  \nonumber \\
&+&J(q^a,q^b,p_0^{ab}) {\cal {Y}}^{ab,cd} p_0^{ab}M_\mu M_\sigma \cdot B_{\sigma\nu}(P^{ab})+\frac{1}{\xi} {\cal {W}}^{ab,cd} p_0^{ab} P_\mu^{ab} M_\nu+J(q^a,q^b,p_0^{ab}) {\cal {Z}}^{ab,cd} p_0^{ab}M_\mu P_\nu^{ab}  \nonumber \\
&& \xlongequal[]{a\neq b}\delta^{ad}\delta^{bc}\delta_{\mu\nu}\, .
\ea
Since the extra projection operator $M_\mu M_\nu$ is only orthogonal to $A_{\mu\nu}(P^{ab})$, as compared to Eq.~(\ref{49}), many new terms associated with $ {\cal {W}}^{ab,cd}$ and $J(q^a,q^b,p_0^{ab})$ are present in the above equation.
Similarly, by requiring the coefficients of all the Lorentz tensor structures expect $\delta_{\mu\nu}$ to vanish, the resummed propagator in Eq.~(\ref{a4}) can be determined by the following result,
\ba\la{repropert2}
{\cal {X}}^{ab,cd}&=&\frac{1}{(P^{ab})^2 + F_T(q^a,q^b,P^{ab})}\, , \nonumber \\
{\cal {Y}}^{ab,cd}&=&\frac{(p^{ab}_0)^2}{(P^{ab})^2}\frac{1+\xi J(q^a,q^b,p_0^{ab})/(P^{ab})^2}{F^0_L(q^a,q^b,P^{ab})+\Big[ (p^{ab}_0)^2 + F_L(q^a,q^b,P^{ab})\Big]\xi J(q^a,q^b,p_0^{ab})/(P^{ab})^2}\, , \nonumber \\
{\cal {Z}}^{ab,cd}&=&\frac{\xi}{(P^{ab})^4}\frac{F^0_L(q^a,q^b,P^{ab})+(p_0^{ab})^2 J(q^a,q^b,p_0^{ab})/(P^{ab})^2}{F^0_L(q^a,q^b,P^{ab})+\Big[ (p^{ab}_0)^2 + F_L(q^a,q^b,P^{ab})\Big]\xi J(q^a,q^b,p_0^{ab})/(P^{ab})^2}\, , \nonumber \\
{\cal {W}}^{ab,cd}&=&-\frac{\xi J(q^a,q^b,p_0^{ab})/(P^{ab})^2}{F^0_L(q^a,q^b,P^{ab})+\Big[ (p^{ab}_0)^2 + F_L(q^a,q^b,P^{ab})\Big]\xi J(q^a,q^b,p_0^{ab})/(P^{ab})^2}\, ,
\ea
where
\be
F^0_L(q^a,q^b,P^{ab})=(P^{ab})^2 + \frac{(P^{ab})^2}{(p^{ab}_0)^2} F_L(q^a,q^b,P^{ab})+\frac{p^2}{(P^{ab})^2} J(q^a,q^b,p_0^{ab})\,.
\ee
Notice that we omit a common color factor $\delta^{ad}\delta^{bc}$ in Eq.~(\ref{repropert2}).

\section{Resummed gluon propagator in $SU(3)$ }
\la{appa}
In this appendix, we will present the calculation of the resummed gluon propagator ${\tilde D}_{\mu\nu}^{a,b}(P)$ in $SU(3)$ under the special constraint
\be\la{cons}
\sum_e {\tilde D}_{\mu\nu}^{e,c}(P)=0\, , \quad\quad (c=1,2,3)\,.
\ee

Starting from Eq.~(\ref{rex}) with $c=1$, we have the following equations
\begin{equation}
\left\{
\begin{aligned}
&{\cal{A}}^{2,2}({\cal {X}}^{2,1}-{\cal {X}}^{1,1})+{\cal{A}}^{2,3}({\cal {X}}^{3,1}-{\cal {X}}^{1,1})=-\frac{1}{3} \\
&{\cal{A}}^{3,3}({\cal {X}}^{3,1}-{\cal {X}}^{1,1})+{\cal{A}}^{3,2}({\cal {X}}^{2,1}-{\cal {X}}^{1,1})=-\frac{1}{3} \\
\end{aligned}
\right.\, .
\end{equation}
The solutions of the above equations can be easily obtained
\begin{equation}\la{so1}
\left\{
\begin{aligned}
&{\cal {X}}^{2,1}-{\cal {X}}^{1,1}=\frac{1}{3} \frac{{\cal{A}}^{2,3}-{\cal{A}}^{3,3}}{{\cal{A}}^{2,2}{\cal{A}}^{3,3}-{\cal{A}}^{2,3}{\cal{A}}^{2,3}}\\
&{\cal {X}}^{3,1}-{\cal {X}}^{1,1}=\frac{1}{3} \frac{{\cal{A}}^{2,3}-{\cal{A}}^{2,2}}{{\cal{A}}^{2,2}{\cal{A}}^{3,3}-{\cal{A}}^{2,3}{\cal{A}}^{2,3}}\\
\end{aligned}
\right.\, .
\end{equation}
In addition, setting $a=1$ in Eq~(\ref{rexin}), the equations for ${\cal {X}}^{1,2}-{\cal {X}}^{1,1}$ and ${\cal {X}}^{1,3}-{\cal {X}}^{1,1}$ read
\begin{equation}
\left\{
\begin{aligned}
&{\cal{A}}^{2,2}({\cal {X}}^{1,2}-{\cal {X}}^{1,1})+{\cal{A}}^{3,2}({\cal {X}}^{1,3}-{\cal {X}}^{1,1})=-\frac{1}{3} \\
&{\cal{A}}^{3,3}({\cal {X}}^{1,3}-{\cal {X}}^{1,1})+{\cal{A}}^{2,3}({\cal {X}}^{1,2}-{\cal {X}}^{1,1})=-\frac{1}{3} \\
\end{aligned}
\right.\, .
\end{equation}
Since ${\cal{A}}^{a,b}={\cal{A}}^{b,a}$, it is obvious to see that ${\cal {X}}^{1,2}-{\cal {X}}^{1,1}={\cal {X}}^{2,1}-{\cal {X}}^{1,1}$ and ${\cal {X}}^{1,3}-{\cal {X}}^{1,1}={\cal {X}}^{3,1}-{\cal {X}}^{1,1}$, namely, ${\cal {X}}^{1,2}={\cal {X}}^{2,1}$, ${\cal {X}}^{1,3}={\cal {X}}^{3,1}$. The solutions for other unknowns ${\cal {X}}^{a,c}-{\cal {X}}^{c,c}$, ${\cal {Y}}^{a,c}-{\cal {Y}}^{c,c}$ can be obtained by simply repeating the above procedure which we don't show here.

Adding the constraint ${\cal {X}}^{1,1}+{\cal {X}}^{2,1}+{\cal {X}}^{3,1}=0$ to Eq.~(\ref{so1}), the solutions for ${\cal {X}}^{a,1}$ with $a=1,2,3$ can be obtained as
\begin{equation}\la{su3sol}
\left\{
\begin{aligned}
&{\cal {X}}^{1,1}=\frac{1}{9} \frac{{\cal{A}}^{1,1}-4{\cal{A}}^{2,3}}{{\cal{A}}^{2,2}{\cal{A}}^{3,3}-{\cal{A}}^{2,3}{\cal{A}}^{2,3}}\\
&{\cal {X}}^{2,1}=-\frac{1}{9} \frac{{\cal{A}}^{2,1}+2{\cal{A}}^{3,3}}{{\cal{A}}^{2,2}{\cal{A}}^{3,3}-{\cal{A}}^{2,3}{\cal{A}}^{2,3}}\\
&{\cal {X}}^{3,1}=-\frac{1}{9} \frac{{\cal{A}}^{3,1}+2{\cal{A}}^{2,2}}{{\cal{A}}^{2,2}{\cal{A}}^{3,3}-{\cal{A}}^{2,3}{\cal{A}}^{2,3}}\\
\end{aligned}
\right.\, .
\end{equation}

The determination of other components of ${\cal{X}}^{a,b}$ follows exactly the same way. For example, impose the constraint ${\cal {X}}^{1,2}+{\cal {X}}^{2,2}+{\cal {X}}^{3,2}=0$ on Eq.~(\ref{rex}) with $c = 2$, then ${\cal {X}}^{a,2}$ with $a = 1,2, 3$ are uniquely determined. In addition, we find that there exists a general expression for ${\cal {X}}^{a,b}$ which reads
\begin{equation}\la{su3sol2}
{\cal {X}}^{a,b}=-2\frac{{\cal P}^{a,b}\sum^\prime_{e,f}{\cal{A}}^{e,f}+{\cal{A}}^{a,b}}{\sum^\prime_{e,f}({\cal{A}}^{e,e}{\cal{A}}^{f,f}-{\cal{A}}^{e,f}{\cal{A}}^{e,f})}\, ,\quad\quad (a,b=1,2,3)\,.
\end{equation}

Given the above discussions, the determination of ${\cal{Y}}^{a,b}$ is straightforward, we only list the result for completeness.
\begin{equation}\la{su3sol3}
{\cal {Y}}^{a,b}=-\frac{2 p_0^4}{P^4}\frac{{\cal P}^{a,b}\sum^\prime_{e,f}{\cal{B}}^{e,f}+{\cal{B}}^{a,b}}{\sum^\prime_{e,f}({\cal{B}}^{e,e}{\cal{B}}^{f,f}-{\cal{B}}^{e,f}{\cal{B}}^{e,f})}\, ,\quad\quad (a,b=1,2,3)\,.
\end{equation}

Finally, the solutions for ${\cal{Z}}^{a,b}$ are unchanged as compared to the bare propagator. For general $SU(N)$, using the constraint $\sum_e {\cal Z}^{e,c}=0$, we have
\be
{\cal{Z}}^{a,b}={\cal P}^{a,b}\frac{\xi}{P^4}\,.
\ee

In terms of ${\tilde{\cal {A}}}^{a,b}$ given in Eq.~(\ref{tildeA}), the final expression for ${\tilde D}_{\mu\nu}^{a,b}(P)$ of $SU(3)$ takes the following form
\ba\la{su3pro}
\tilde{D}^{a,b}_{\mu\nu}(P)&=&\frac{1}{{\tilde P}^2}\frac{{\cal P}^{a,b}-2 \big(\frac{T^2}{{\tilde T}^2}\big)\big(1-\frac{P^2}{{\tilde P}^2}\big)\big(\sum^\prime_{ef}{\tilde {\cal A}}^{e,f}{\cal P}^{a,b}+{\tilde {\cal A}}^{a,b}\big)}{1-2\big(\frac{T^2}{{\tilde T}^2}\big) \big(1-\frac{P^2}{{\tilde P}^2}\big)\sum^\prime_{ef}{\tilde {\cal A}}^{e,f}+6\big(\frac{T^2}{{\tilde T}^2}\big)^2\big(1-\frac{P^2}{{\tilde P}^2}\big)^2 \sum^\prime_{efg}{\tilde {\cal A}}^{e,f}{\tilde {\cal A
}}^{g,e}}A_{\mu\nu}(P)\nonumber\\
&+&\frac{p_0^4/P^4}{{\tilde p_0}^2}\frac{{\cal P}^{a,b}-2\big(\frac{T^2}{{\tilde T}^2}\big)\big(1-\frac{p_0^2}{{\tilde p_0}^2}\big)\big(\sum^\prime_{ef}{\tilde {\cal A}}^{e,f}{\cal P}^{a,b}+{\tilde {\cal A}}^{a,b}\big)}{1- 2\big(\frac{T^2}{{\tilde T}^2}\big)\big(1-\frac{p_0^2}{{\tilde p_0}^2}\big)\sum^\prime_{ef}{\tilde {\cal A}}^{e,f}+6\big(\frac{T^2}{{\tilde T}^2}\big)^2\big(1-\frac{p_0^2}{{\tilde p_0}^2}\big)^2  \sum^\prime_{efg}{\tilde {\cal A}}^{e,f}{\tilde {\cal A
}}^{g,e}}B_{\mu\nu}(P)+\xi\frac{{\cal P}^{a,b}}{P^4}P_\mu P_{\nu}\, ,\nonumber\\
\ea
were ${\tilde T}$ is defined in Eq.~(\ref{newp}).

For zero background field, due to the vanishing ${\tilde{\cal {A}}}^{a,b}$, the diagonal gluon propagator becomes
\be
\tilde{D}^{a,b}_{\mu\nu}(P, q\rightarrow0)=\bigg(\frac{1}{{\tilde P}^2}A_{\mu\nu}(P)+\frac{p_0^4/P^4}{{\tilde p}_0^2}B_{\mu\nu}(P)+\frac{\xi}{ P^4}P_\mu P_{\nu}\bigg){\cal P}^{a,b}\, .
\ee

We point out that with Eqs.~(\ref{didef}) and (\ref{cons}), all the diagonal gluon propagators for $SU(3)$ are uniquely determined without resorting to Eq.~(\ref{ddinre}). The obtained solutions for ${\tilde D}_{\mu\nu}^{a,b}$ is symmetric in color space, {\em i.e.}, ${\tilde D}_{\mu\nu}^{a,b}={\tilde D}_{\mu\nu}^{b,a}$, as a result, Eq.~(\ref{ddinre}) is satisfied automatically. In fact, we find that such a conclusion actually holds for general $SU(N)$ if the special constraint $\sum_e {\tilde D}_{\mu\nu}^{e,c}(P)=0$ with $c=1,2,\cdots\,N$ is adopted.

\section{Calculation of $\sum_{a,b}{\cal P}^{a,b} {\tilde{D}}_{\mu\nu}^{a,b}(P)$ for general $SU(N)$}
\la{appb}

To make our presentation compact, we first introduce the following short-hand notations. The $N \times N$ matrix ${\cal {A}}$ in color space has the explicit form
\begin{equation}\la{mx1}
{\cal {A}}=
\begin{pmatrix}
{\cal {A}}^{1,1} & {\cal {A}}^{1,2}  & \cdots   & {\cal {A}}^{1,N}   \\
{\cal {A}}^{2,1} & {\cal {A}}^{2,2}  & \cdots   & {\cal {A}}^{2,N}  \\
\vdots & \vdots  & \ddots   & \vdots  \\
{\cal {A}}^{N,1} & {\cal {A}}^{N,2}  & \cdots  & {\cal {A}}^{N,N}
\end{pmatrix}
\, ,
\end{equation}
with
\be\la{mA}
{\cal {A}}^{a,b}=
\left\{
\begin{aligned}
&{\tilde P}^2-\frac{1}{N}{\tilde P}^2 + 2g^2T^2\Pi_T(P)\sum_{e}{\hat B}_2(q^{a e})\quad\quad {\rm for}\quad a=b \\
&-\frac{1}{N}{\tilde P}^2 -2g^2T^2\Pi_T(P){\hat B}_2(q^{a b})\quad\quad\quad\quad\,\,\,\,\,\,{\rm for}\quad a\neq b \\
\end{aligned}
\right.
\, ,
\ee
where the definitions of ${\tilde P}^2$ and ${\hat B}_2(x)$ can be found in Eqs.~(\ref{newp}) and (\ref{tildeA}), respectively. In addition, ${\cal {A}}^{[a]}$ is defined as a $(N-1) \times (N-1)$ matrix which is obtained by removing the $2N-1$ elements in the $a^{{\rm th}}$ row and $a^{{\rm th}}$ column from ${\cal {A}}$. Because $\sum_e {\cal {A}}^{e,c}=\sum_e {\cal {A}}^{c,e}=0$, the determinant of ${\cal {A}}$ vanishes. For the same reason, one can also show that the determinant of ${\cal {A}}^{[a]}$ is independent on the value of $a$ where $a=1,2,\cdots,n$. Furthermore, ${\cal {A}}^{[a]}\{b\}$ is used to denote a matrix that constructed by two successive steps. First, we replace the $N$ elements in column $b$ of matrix ${\cal {A}}$ with $-1/N$, then, remove the $2N-1$ elements in the $a^{{\rm th}}$ row and $a^{{\rm th}}$ column from the previous obtained matrix.

We start by considering the solutions for ${\cal {X}}^{a,c}+{\cal {X}}^{c,a}-{\cal {X}}^{a,a}-{\cal {X}}^{c,c}$ for general $SU(N)$.  According to Eq.~(\ref{rex}), we choose $c$ to be some fixed value $j$ and solve this equation for the $N-1$ unknowns ${\cal {X}}^{a,j}-{\cal {X}}^{j,j}$ with $a\neq j$. Using the Cramer's rule, the solution for one specified unknown ${\cal {X}}^{i,j}-{\cal {X}}^{j,j}$ is formally written as
\be
{\cal {X}}^{i,j}-{\cal {X}}^{j,j}=\frac{\big|{\cal {A}}^{[j]}\{i\}\big|}{\big|{\cal {A}}^{[j]}\big|}\, .
\ee
Similarly, we choose $a$ to be some fixed value $i$ in Eq.~(\ref{rexin}) and the solution for ${\cal {X}}^{j,i}-{\cal {X}}^{i,i}$ reads
\be
{\cal {X}}^{j,i}-{\cal {X}}^{i,i}=\frac{\big|{\cal {A}}^{[i]}\{j\}\big|}{\big|{\cal {A}}^{[j]}\big|}\, .
\ee
Summing up the above two equations, we have the following expression for ${\cal {X}}^{i,j}+{\cal {X}}^{j,i}-{\cal {X}}^{i,i}-{\cal {X}}^{j,j}$
\be\la{re}
{\cal {X}}^{i,j}+{\cal {X}}^{j,i}-{\cal {X}}^{i,i}-{\cal {X}}^{j,j}=-\frac{\big|{\cal {A}}^{[i,j]}\big|}{\big|{\cal {A}}^{[i]}\big|}=-\frac{\big|{\cal {A}}^{[i,j]}\big|}{\big|{\cal {A}}^{[j]}\big|}\, ,
\ee
where we have used
\be\la{detre}
\big|{\cal {A}}^{[j]}\{i\}\big|+\big|{\cal {A}}^{[i]}\{j\}\big|=-\big|{\cal {A}}^{[i,j]}\big|\, ,
\ee
and ${\cal {A}}^{[i,j]}$ is a $(N-2)\times (N-2)$ matrix obtained by removing the $4N-4$ elements in the $i^{{\rm th}}$ and $j^{{\rm th}}$ rows as well as the $i^{{\rm th}}$ and $j^{{\rm th}}$ columns from matrix ${\cal {A}}$.

Although it is not very obvious, Eq.~(\ref{detre}) can be straightforwardly obtained in the following way. Performing the sequential elementary row and column operations\footnote{$R_a \leftrightarrow R_b$ stands for swapping rows $a$ and $b$. The column operation $C_a \leftrightarrow C_b$ is for swapping columns $a$ and $b$.} on ${\cal {A}}^{[i]}\{j\}$, $R_j \leftrightarrow R_{j-1},R_{j-1} \leftrightarrow R_{j-2},\cdots R_2 \leftrightarrow R_1$, then $C_j \leftrightarrow C_{j-1},C_{j-1} \leftrightarrow C_{j-2},\cdots C_2 \leftrightarrow C_1$, the obtained matrix $\underline{{\cal {A}}}^{[i]}\{j\}$ has the same determinant as ${\cal {A}}^{[i]}\{j\}$. After similar transformations, $R_i \leftrightarrow R_{i-1},R_{i-1} \leftrightarrow R_{i-2},\cdots R_2 \leftrightarrow R_1$, then $C_i \leftrightarrow C_{i-1},C_{i-1} \leftrightarrow C_{i-2},\cdots C_2 \leftrightarrow C_1$, ${\cal {A}}^{[j]}\{i\}$ becomes $\underline{{\cal {A}}}^{[j]}\{i\}$ while the determinant also keeps unchanged. These two matrices $\underline{{\cal {A}}}^{[i]}\{j\}$ and $\underline{{\cal {A}}}^{[j]}\{i\}$ are identical except the elements in the first row. Therefore, we arrive at the following equation
\be
\big|{\cal {A}}^{[j]}\{i\}\big|+\big|{\cal {A}}^{[i]}\{j\}\big|=\big|\underline{{\cal {A}}}^{[j]}\{i\}\big|+\big|\underline{{\cal {A}}}^{[i]}\{j\}\big|\equiv\big|{\cal {A}}_{\rm sum}\big|\, .
\ee
The elements in the first row of the introduced $(N-1)\times (N-1)$ matrix ${\cal {A}}_{\rm sum}$ are given by
\be
\big({\cal {A}}_{\rm sum}\big)^{1,a}=\big(\underline{{\cal {A}}}^{[j]}\{i\}\big)^{1,a}+\big(\underline{{\cal {A}}}^{[i]}\{j\}\big)^{1,a}\, ,
\ee
with $a=1,2,\cdots, N-1$ while other elements are the same as $\underline{{\cal {A}}}^{[j]}\{i\}$ or $\underline{{\cal {A}}}^{[i]}\{j\}$. Adding rows $2$ to $N-1$ to the first row, $\sum_{i=1}^{N-1} R_i\rightarrow R_1$, such an elementary row operation doesn't change the determinant of ${\cal {A}}_{\rm {sum}}$ and the resulting matrix is denoted as $\underline{{\cal {A}}}_{\rm sum}$. On the one hand, due to $\sum_e {\cal {A}}^{e,c}=\sum_e {\cal {A}}^{c,e}=0$, the only non-vanishing element in the first row of $\underline{{\cal {A}}}_{\rm sum}$ is $\big(\underline{{\cal {A}}}_{\rm sum}\big)^{1,1}=-1$. On the other, after removing the elements in the first row and column from $\underline{{\cal {A}}}_{\rm sum}$, the resulting $(N-2)\times (N-2)$ matrix is nothing but ${\cal {A}}^{[i,j]}$. Then it is clear to see the validity of Eq.~(\ref{detre}).

The determinants in Eq.~(\ref{re}) is not easy to compute for arbitrary $N$ which depend on the momentum $P$ as well as the background field $A_0^{\rm cl}$. In this work, we are particularly interested in the influence of the background field on the resummed gluon propagators, therefore, it makes sense to eliminate the $P$-dependence in the determinants which as a result will depend only on $A_0^{\rm cl}$. We find this is doable with the following two steps.

As shown in Eq.~(\ref{mA}), there is a common term ${\tilde P}^2$ appearing in $\big({\cal {A}}^{[i,j]}\big)^{a,a}$. With the
basic properties of the determinant of a matrix, the first step is to rewrite $\big|{\cal {A}}^{[i,j]}\big|$ as
\be
\big|{\cal {A}}^{[i,j]}\big|=\sum_{k=0}^{N-2}{\tilde P}^{2(N-2-k)} {\hat S}^{[i,j]}_k\, .
\ee
In the above equation, ${\hat S}^{[i,j]}_k$ denotes a sum of the determinants which reads
\be
{\hat S}^{[i,j]}_k = {\sum_{a_{\{k\}}}}^{[i,j]} \begin{vmatrix}
{\hat{{\cal{A}}}}^{a_1,a_1} & {\hat{{\cal{A}}}}^{a_1,a_2}  & \cdots   & {\hat{{\cal{A}}}}^{a_1,a_k}   \\
{\hat{{\cal{A}}}}^{a_2,a_1} & {\hat{{\cal{A}}}}^{a_2,a_2}  & \cdots   & {\hat{{\cal{A}}}}^{a_2,a_k}  \\
\vdots & \vdots  & \ddots   & \vdots  \\
{\hat{{\cal{A}}}}^{a_k,a_1} & {\hat{{\cal{A}}}}^{a_k,a_2}  & \cdots  & {\hat{{\cal{A}}}}^{a_k,a_k}
\end{vmatrix}\, ,
\ee
whit the special case ${\hat S}^{[i,j]}_0=1$. The shorthand notation $a_{\{k\}}$ has been defined in Sec.~\ref{red}. In addition, ${\sum}^{[i,j]}$ requires that summation indices $a_1,a_2,\cdots, a_k$ can not be equal to the specified values $i$ or $j$ when run from $1$ to $N$. The $k\times k$ matrix ${\hat{{\cal{A}}}}$ in the above equation is given by
\be
{\hat{{\cal{A}}}}^{a,b}={\cal A}^{a,b}-\delta^{ab}{\tilde P}^2\,.
\ee

According to Eq.~(\ref{pd}), in order to compute the quantity ${\cal P}^{ab,cd}{\tilde {D}}^{ab,cd}$, we should actually consider the sum $\sum_{i > j}\big|{\cal{A}}^{[i,j]}\big|$ which can be written as
\be\la{s1}
\sum_{i>j}\big|{\cal{A}}^{[i,j]}\big|=\sum_{k=0}^{N-2}{\tilde P}^{2(N-2-k)}{\rm C}^{\,2}_{N-k} {\hat S}_k\, ,
\ee
where ${\rm C}^{\,2}_{N-k}$ is the binomial coefficient and ${\hat S}_k$ is differentiated from ${\hat S}^{[i,j]}_k$ only by the fact that the set of indices $a_{\{k\}}$ now run from $1$ to $N$. Accordingly, the upper-script of ${\hat S}^{[i,j]}_k$ has been removed. When $k=0$, we have ${\hat S}_0=1$.

Notice that every element in matrix ${\hat{{\cal{A}}}}$ has a term $-{\tilde P}^2/N$. The second step is to take this term out from ${\hat S}_k$ and we can show
\ba
 \begin{vmatrix}
{\hat{{\cal{A}}}}^{a_1,a_1} & {\hat{{\cal{A}}}}^{a_1,a_2}  & \cdots   & {\hat{{\cal{A}}}}^{a_1,a_k}   \\
{\hat{{\cal{A}}}}^{a_2,a_1} & {\hat{{\cal{A}}}}^{a_2,a_2}  & \cdots   & {\hat{{\cal{A}}}}^{a_2,a_k}  \\
\vdots & \vdots  & \ddots   & \vdots  \\
{\hat{{\cal{A}}}}^{a_k,a_1} & {\hat{{\cal{A}}}}^{a_k,a_2}  & \cdots  & {\hat{{\cal{A}}}}^{a_k,a_k}
\end{vmatrix}
&=&-\frac{1}{N}{\tilde P}^2
\sum_{i=1}^k
\begin{vmatrix}
{\bar{\cal {A}}}^{a_1,a_1} & {\bar{\cal {A}}}^{a_1,a_2}  & \cdots   & {\bar{\cal {A}}}^{a_1,a_k}   \\
{\bar{\cal {A}}}^{a_2,a_1} & {\bar{\cal {A}}}^{a_2,a_2}  & \cdots   & {\bar{\cal {A}}}^{a_2,a_k}  \\
\vdots & \vdots  & \ddots   & \vdots  \\
{\bar{\cal {A}}}^{a_k,a_1} & {\bar{\cal {A}}}^{a_k,a_2}  & \cdots  & {\bar{\cal {A}}}^{a_k,a_k}
\end{vmatrix}_{C_i\rightarrow 1}\nonumber\\
&+&
\begin{vmatrix}
{\bar{\cal {A}}}^{a_1,a_1} & {\bar{\cal {A}}}^{a_1,a_2}  & \cdots   & {\bar{\cal {A}}}^{a_1,a_k}   \\
{\bar{\cal {A}}}^{a_2,a_1} & {\bar{\cal {A}}}^{a_2,a_2}  & \cdots   & {\bar{\cal {A}}}^{a_2,a_k}  \\
\vdots & \vdots  & \ddots   & \vdots  \\
{\bar{\cal {A}}}^{a_k,a_1} & {\bar{\cal {A}}}^{a_k,a_2}  & \cdots  & {\bar{\cal {A}}}^{a_k,a_k}
\end{vmatrix}
\, ,
\ea
where $C_i\rightarrow 1$ indicates the replacement of all the elements in column $i$ with $1$ and the matrix ${\bar{\cal {A}}}^{a,b}$ is defined as
\be
{\bar{\cal {A}}}^{a,b}
= 2g^2T^2\Pi_T(P)\Big( \sum_{e} {\hat B}_2(q^{a e})\delta^{ab}-{\hat B}_2(q^{a b})(1-\delta^{ab})\Big)\, .
\ee
Because ${\bar{\cal {A}}}^{b,a}={\bar{\cal {A}}}^{a,b}$ and $\sum_e{\bar{\cal {A}}}^{e,c}=0$, we can derive the following identity
\ba
\sum_{a_{\{k\}}}\sum_{i=1}^k
\begin{vmatrix}
{\bar{\cal {A}}}^{a_1,a_1} & {\bar{\cal {A}}}^{a_1,a_2}  & \cdots   & {\bar{\cal {A}}}^{a_1,a_k}   \\
{\bar{\cal {A}}}^{a_2,a_1} & {\bar{\cal {A}}}^{a_2,a_2}  & \cdots   & {\bar{\cal {A}}}^{a_2,a_k}  \\
\vdots & \vdots  & \ddots   & \vdots  \\
{\bar{\cal {A}}}^{a_k,a_1} & {\bar{\cal {A}}}^{a_k,a_2}  & \cdots  & {\bar{\cal {A}}}^{a_k,a_k}
\end{vmatrix}_{C_i\rightarrow 1}
=N\sum_{a_{\{k-1\}}}
\begin{vmatrix}
{\bar{\cal {A}}}^{a_1,a_1} & {\bar{\cal {A}}}^{a_1,a_2}  & \cdots   & {\bar{\cal {A}}}^{a_1,a_{k-1}}   \\
{\bar{\cal {A}}}^{a_2,a_1} & {\bar{\cal {A}}}^{a_2,a_2}  & \cdots   & {\bar{\cal {A}}}^{a_2,a_{k-1}}  \\
\vdots & \vdots  & \ddots   & \vdots  \\
{\bar{\cal {A}}}^{a_{k-1},a_1} & {\bar{\cal {A}}}^{a_{k-1},a_2}  & \cdots  & {\bar{\cal {A}}}^{a_{k-1},a_{k-1}}
\end{vmatrix}\, ,\nonumber \\
\ea
which leads to our final result for the determinant of ${\cal A}^{[i,j]}$
\be\la{nu}
\sum_{i>j}\big|{\cal{A}}^{[i,j]}\big|={\tilde P}^{2(N-2)}\sum_{k=0}^{N-2} \Big(\frac{6 }{N} \Big)^k \Big(\frac{T^2}{{\tilde T}^2}\Big)^k\Big(1-\frac{P^2}{\tilde{P}^2}\Big)^k(N-k-1){\tilde S}_k\, ,
\ee
where ${\tilde T}^2$ is defined in Eq.~(\ref{newp}). According to Eq.~(\ref{tildeS}), ${\tilde S}_k$ denotes a sum of determinants of matrix ${\tilde{\cal {A}}}$ with ${\tilde{\cal {A}}}^{a,b} = {\bar{\cal {A}}}^{a,b}/(2g^2T^2\Pi_T(P)) $.The matrix element ${\tilde{\cal {A}}}^{a,b}$ depends only on the background field and vanishes when $A_0^{\rm cl}=0$. As before, ${\tilde S}_0=1$, therefore, only the term $k=0$ contributes for vanishing $A_0^{\rm cl}$.

The corresponding calculation of the determinant of ${\cal {A}}^{[j]}$ in Eq.~(\ref{re}) can be carried out in a similar way. As mentioned before, the determinants of ${\cal {A}}^{[a]}$ for $a=1,2,\cdots,N$ are all equal, so we have
\ba
\big|{\cal {A}}^{[j]}\big|&=&\frac{1}{N} \sum_{e=1}^{N} \big|{\cal {A}}^{[e]}\big|=\sum_{k=0}^{N-1}{\tilde P}^{2(N-1-k)}\frac{N-k}{N} {\hat S}_k\, \nonumber\\
&=&\frac{{\tilde P}^{2(N-1)}}{N}\sum_{k=0}^{N-1} \Big(\frac{6 }{N} \Big)^k \Big(\frac{T^2}{{\tilde T}^2}\Big)^k\Big(1-\frac{P^2}{\tilde{P}^2}\Big)^k{\tilde S}_k\,.
\ea

Given the above discussions, the calculation of ${\cal {Y}}^{i,j}+{\cal {Y}}^{j,i}-{\cal {Y}}^{i,i}-{\cal {Y}}^{j,j}$ becomes a trivial repetition. After taking into account Eq.~(\ref{recaly}), it is straightforward to write down the final expression for $\sum_{a,b}{\cal P}^{a,b} {\tilde{D}}_{\mu\nu}^{a,b}(P)$ which has been given in Eq.~(\ref{udd}).

\end{document}